\documentclass[aoas]{imsart}

\RequirePackage{amsthm,amsmath,amsfonts,amssymb}
\RequirePackage[authoryear]{natbib}
\RequirePackage[colorlinks,citecolor=blue,urlcolor=blue]{hyperref}
\RequirePackage{graphicx}

\usepackage{bbold}
\usepackage[utf8]{inputenc}
\usepackage{amsmath,amsfonts}
\usepackage{multirow}
\usepackage{caption,subcaption}
\usepackage{booktabs}

\startlocaldefs
\theoremstyle{plain}
\newtheorem{theorem}{Theorem}
\DeclareMathOperator*{\argmin}{arg\,min}

\endlocaldefs

\begin{document}

\begin{frontmatter}
\title{Topological Learning for Brain Networks}
\runtitle{Topological Learning for Brain Networks}

\begin{aug}
\author{\fnms{Tananun}~\snm{Songdechakraiwut}\ead[label=e1]{songdechakra@wisc.edu}}
\and
\author{\fnms{Moo K.}~\snm{Chung}\ead[label=e2]{mkchung@wisc.edu}}
\address{Department of Biostatistics and Medical Informatics\\
University of Wisconsin--Madison\\
Madison, WI 53705\\
United States\\
\printead{e1,e2}}

\end{aug}

\begin{abstract}
This paper proposes a novel topological learning framework that integrates networks of different sizes and topology through persistent homology. Such challenging task is made possible through the introduction of a computationally efficient topological loss. The use of the proposed loss bypasses the intrinsic computational bottleneck associated with matching networks. We validate the method in extensive statistical simulations to assess its effectiveness when discriminating networks with different topology. The method is further demonstrated in a twin brain imaging study where we determine if brain networks are genetically heritable. The challenge here is due to the difficulty of overlaying the topologically different functional brain networks obtained from resting-state functional MRI onto the template structural brain network obtained through diffusion MRI.
\end{abstract}


\end{frontmatter}


\section{Introduction}

Network analysis has experienced tremendous advances in the last few decades. Networks are often represented using graphs consisting of  nodes and edges. In standard network analysis, the focus is mainly on analyzing how data at nodes are interacting with each other. Strength of the interaction is represented as edge weights. In social science, human interactions are modeled by viewing humans as nodes and human interaction as edges \citep{scott.1988}. In molecular studies, interatomic distances in a molecule are measured across atoms and serve as edge weights while the atoms themselves serve as nodes \citep{chung.2021.MICCAI,xia2014persistent}. In brain imaging studies, the whole brain is parcellated into  hundreds of disjoint regions, which then serve as nodes, while brain activities between parcellations, based on correlations, serve as edge weights \citep{arslan.2018,desikan.2006,fornito.2016,hagmann.2007,tzourio.2002}.

In graph theory based network analyses \citep{sporns.2003, vanwijk.2010},  graph theory features such as node degrees and clustering coefficients are often obtained from adjacency matrices after thresholding edge weights. The final statistical results may be different depending on the threshold choice \citep{lee2012persistent}. This motivates the development of a multiscale network model that provides consistent results and interpretation regardless of the choice of thresholding. Topological Data Analysis (TDA) \citep{edelsbrunner2000topological,wasserman2018topological}, a general framework based on algebraic topology, can provide a novel solution to the multiscale network analysis challenge. Instead of examining networks using graphs at one fixed scale, the persistent homology technique in TDA identifies persistent topological features that are robust across scales. 

Numerous TDA studies have been applied to increasingly diverse biomedical problems such as genetics \citep{chung2017exact,chung2019exact},
epileptic seizure detection \citep{wang2018topological}, sexual dimorphism in the human brain \citep{songdechakraiwut2020dynamic}, analysis of brain arteries \citep{bendich2016persistent}, image segmentation \citep{clough2019explicit}, image classification  \citep{chen2019topological,reininghaus2015stable,singh2014topological}, clinical predictive model \citep{crawford2020predicting} and persistence-based clustering \citep{chazal2013persistence}.

Persistent homology has emerged as a powerful mathematical representation for understanding, characterizing and quantifying topology of networks. In persistent homology, topological features are measured across different spatial resolutions.  As the resolution changes, such features are born and die. Persistent homology associates the life-time to these features in the form of 1D intervals from birth to death. The collection of such intervals is summarized as a {\em barcode}, which characterizes the topology of underlying data \citep{ghrist2008barcodes}. Long-lived barcodes persist over a long range of resolutions and are considered as signal  \citep{carlsson2009topology}. Recent work proposed {\em topological loss} that penalizes barcodes for image segmentation problem \citep{hu2019topology}. While this approach allows to incorporate topological information into the segmentation problem, the method has been limited to image segmentation with a small number of topological features due to its expensive optimization process involving $O(|V|^6)$ run-time with $|V|$ number of nodes \citep{edmonds1972theoretical,kerber2017geometry}. Barcodes are typically computed at a finite set of pre-specified resolutions. A sufficient number of such resolutions is required to give a reasonably accurate estimation of barcodes, which quickly increases computational complexity as the size of data increases \citep{chung2019statistical,hu2019topology}. This is impractical in brain networks with far larger number of topological features involving hundreds of connected components and thousands of cycles. In this paper, we propose a more principled approach that learns the topological structure of brain networks with large number of topological features in $O(|E| \log |V|)$ run-time with $|E|$ number of edges and $|V|$ number of nodes. The proposed method bypasses the intrinsic computational bottleneck and thus enables us to perform various topology computations and optimizations at every resolution.

We illustrate the proposed method using the resting-state functional MRI (fMRI) of 194 twin pairs from the Human Connectome Project (HCP) \citep{van2012human,van2013wu}. HCP twin brain imaging data is considered as the {\em gold standard}, where the zigosity is confirmed by the blood and saliva test \citep{gritsenko.2020}. Monozygotic (MZ) twins share 100\% of genes while  dizygotic (DZ) twins share 50\% of genes \citep{falconer.1995}.  MZ-twins are more similar or concordant than DZ-twins for cognitive aging, cognitive dysfunction and Alzheimer's disease \citep{reynolds.2015}. These genetic differences allow us to pull apart and examine genetic and environmental influences easily {\em in vivo}. The difference between MZ- and DZ-twins quantify the extent to which phenotypes are influenced by genetic factors. If MZ-twins show more similarity on a given trait compared to DZ-twins, this provides evidence that genes significantly influence that trait. Previous twin brain imaging studies mainly used univariate imaging phenotypes such as brain cortical thickness \citep{mckay.2014}, fractional anisotropy \citep{chiang.2011} and functional activation \citep{blokland.2011,glahn.2010,smit.2008} when determining heritability in a few regions of interest. Compared to prior studies on univariate imaging phenotypes, there are not many  studies on the heritability of the whole brain functional networks \citep{blokland.2011}. Measures of network topology may be worth investigating as intermediate phenotypes that indicate the genetic risk for a neuropsychiatric disorder \citep{bullmore2009complex}. However, the brain network analysis has not yet been adapted for this purpose beyond a small number of regions. Determining the extent of heritability of the whole brain networks is the first necessary prerequisite for identifying network-based endophenotypes.
We employ our method to determine the heritability of brain networks constructed by topologically overlaying functional brain networks from fMRI onto the template structural brain network from diffusion MRI (dMRI). Our method is demonstrated to increase the sensitivity to subtle genetic signals.

\section{Methods}

\subsection{Preliminary}
Consider a complete network represented as a graph $G=(V,w)$ comprising a set of nodes $V$ and unique positive symmetric edge weights $w=(w_{ij})$ satisfying $w_{ij} = w_{ji}$. Note that the method proposed in this paper is {\em translation invariant} so any negative edge weights can be made positive by translations. In addition, the condition of having unique edge weights is not restrictive in practice. Assuming edge weights follow some continuous distribution, the probability of any two edge weights being equal is zero. This is particularly true in functional brain networks where edge weights are given as the Pearson correlation between time series of brain activity \citep{fornito.2016}. In the case of equal edge weights, we can add infinitesimal noise and break the tie. Since the proposed method is based on the Wasserstein distance, which enjoys the stability theorem \citep{cohen2010lipschitz}, adding the infinitesimal noise will not affect the final numerical outcome.

 The cardinality of sets is denoted using $| \cdot |$. The number of nodes and edges are then denoted as $|V|$ and $|E|$. Since $G$ is a complete graph, we have $| E | =  |V| (|V| -1)/2$. Any incomplete graph can be treated as a special case of a complete graph with zero weights. Then we can add infinitesimal noise to those zero weights to break ties. We emphasize that the proposed method works for arbitrary graphs. However, assuming the graph to be complete with unique positive edge weights makes exposition of the method straightforward.

\begin{figure}[t]
\includegraphics[width=1\linewidth]{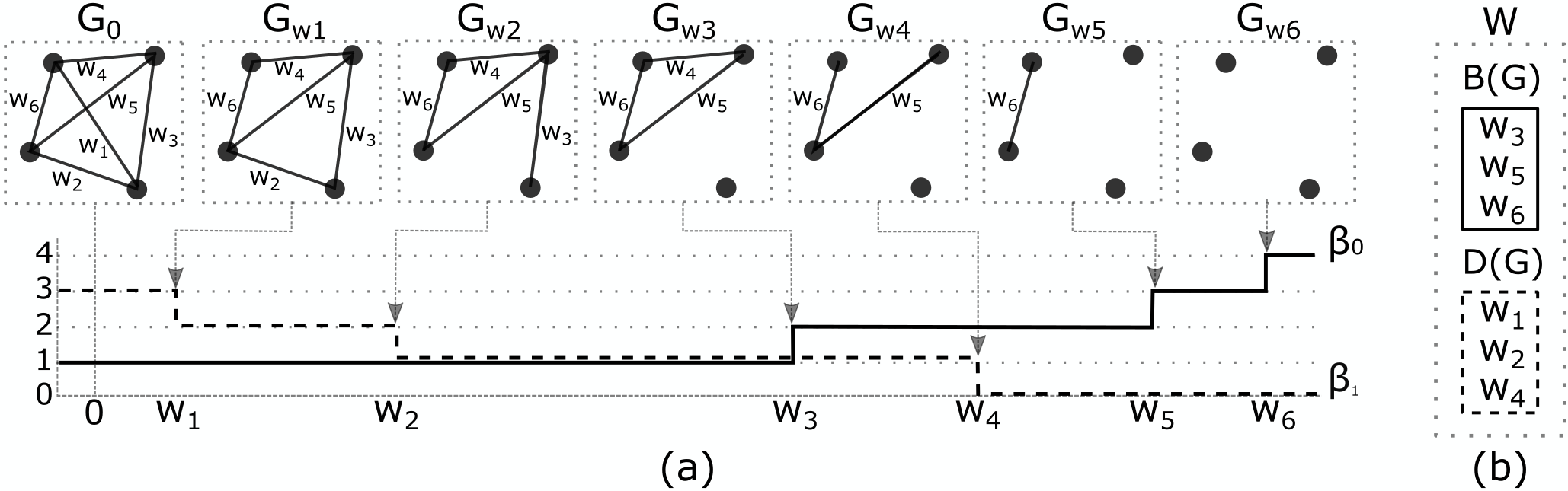}
\centering
\caption{(a) Graph filtration on a four-node network $G$. $\beta_0$ is monotonically increasing while $\beta_1$ is monotonically decreasing over the graph filtration. Connected components are born at the edge weights $w_3,w_5,w_6$ while cycles die at the edge weights $w_1,w_2,w_4$. A cycle consisting of $w_4, w_5, w_6$ persists longer than any other 1D features and is considered as topological signal. 0D barcode $\{ (-\infty,\infty),(w_3,\infty), (w_5,\infty), (w_6,\infty) \}$ is represented using the birth values as  $I_0(G)=\{w_3,w_5,w_6\}$.  1D barcode $\{ (-\infty,w_1), (-\infty,w_2), (-\infty,w_4) \}$ is represented using the death values as $I_1(G)=\{w_1,w_2,w_4\}$. (b)  We can show that 0D and 1D barcodes  uniquely partition the edge weight set as $W  = I_0(G) \cup I_1(G)$ with $I_0(G) \cap l_1(G) = \emptyset$.}
\label{fig:filtration}
\end{figure}

The binary graph $G_\epsilon=(V,w_\epsilon)$ of $G$ is defined as a graph consisting of node set $V$ and binary edge weight $w_\epsilon$ given by 
\[ w_\epsilon = (w_{\epsilon, ij}) =   \begin{cases}
1 &\; \mbox{  if } w_{ij} > \epsilon;\\
0 & \; \mbox{ otherwise}.
\end{cases}
\]
A {\em graph filtration} of $G$ is defined as a collection of nested binary networks \citep{lee2012persistent}:
\[G_{\epsilon_0} \supset G_{\epsilon_1} \supset \cdots \supset G_{\epsilon_k} ,\]
where $\epsilon_0 < \epsilon_1 < \cdots < \epsilon_k$ are called filtration values.
Figure \ref{fig:filtration} displays an example of the graph filtration on a four-node network.

In persistent homology, a 0-dimensional topological feature is a {\em connected component}, which is the set of nodes and edges that are connected to each other by paths. In Figure \ref{fig:filtration}, there is only one connected component in graph $G_0$ since all nodes and edges are connected through paths, while $G_{w6}$ has four connected components since each of the four nodes cannot be reached from the others by any path.  The number of connected components is called the 0-th Betti number $\beta_0$ and we write $\beta_0(G_{0}) = 1$ and $\beta_0(G_{w6}) = 4$.

The 1-dimensional topological feature is a loop or {\em cycle}, which is a path that starts and ends at the same node but no other nodes in the path are overlapping.  In $G_{w3}$, there is one cycle consisting of edges $w_4$, $w_5$ and $w_6$.  The cycle can be algebraically represented as $[w_4] + [w_5] + [w_6]$ with the convention of putting clockwise orientation along the edges.  
In $G_{w_1}$, there are three cycles consisting of $[w_4] +  [w_5]  +  [w_6]$, $-[w_5] + [w_3] + [w_2]$ and $[w_4] + [w_3] +  [w_2] + [w_6]$. However, they are linearly dependent in a sense that the cycle consisting of four nodes can be written as the sum of the two other smaller cycles:
$$[w_4] + [w_3] +  [w_2] + [w_6] = ([w_4] +  [w_5]  +  [w_6]) + ( -[w_5] + [w_3] + [w_2]).$$
Thus, there are only two algebraically independent cycles in $G_{w_1}$. The total number of independent cycles is the 1-st Betti number $\beta_1$. We write  $\beta_1(G_{w3}) = 1$ and $\beta_1(G_{w_1}) = 2$.
During the graph filtration, $\beta_0$ is monotonically increasing while $\beta_1$ is monotonically decreasing \citep{chung2019exact}.
Unlike Rips complexes \citep{ghrist2008barcodes}, which are often used in persistent homology, there are no more higher dimensional topological features than 0D and 1D. 

Persistent homology keeps track of appearances (birth) and disappearances (death) of connected components and cycles over the filtration as well as their {\em persistence} (the duration from birth to death). Longer persistence indicates the presence of larger topological signal  \citep{edelsbrunner2008persistent}. The persistence of topological features are algebraically represented as the collection of intervals $(\epsilon_b,\epsilon_d)$, where a feature appears at the filtration value $\epsilon_b$ and vanishes at the filtration value $\epsilon_d$. The collection of such intervals is called a barcode \citep{adler.2010,ghrist2008barcodes,song.2021.MICCAI}.

\subsection{Birth-death decomposition}
During the graph filtration, once a connected component is born, it never dies. Thus, every death value of connected components is considered as $\infty$. We can safely discard $\infty$ and characterize the 0D barcode for connected components using birth values 
\[I_0(G): \quad \epsilon_{b1} < \epsilon_{b2} < \cdots < \epsilon_{bm_0},\]
where $m_0  = \beta_0(G_{\infty}) -1 = |V|-1.$
$I_0(G)$ forms the maximum spanning tree (MST) of $G$ \citep{lee2012persistent}. 
Cycles in the graph filtration are all considered born at $-\infty$. Thus, we can also discard $-\infty$ and characterize the 1D barcode for cycles using death values 
\[I_1(G): \quad \epsilon_{d1} <  \epsilon_{d2} < \cdots < \epsilon_{d m_1},\]
where $m_1= \beta_1 (G_0) =  (|V| -1) (|V| - 2)/2.$
Removing edge $w_{ij}$ in the graph filtration results in either the birth of a connected component or the death of a cycle. The birth of a connected component and the death of a cycle cannot possibly happen at the same time. Thus, every edge weight must be in either 0D barcode or 1D barcode, but not both (Figure \ref{fig:filtration}-b).

\begin{theorem} 
The birth set $I_0(G)$ and death set $I_1(G)$ partition the edge weight set $W$ such that $W  = I_0(G) \cup I_1(G)$ with $I_0(G) \cap l_1(G) = \emptyset.$
The cardinalities of $I_0(G)$ and $I_1(G)$ are $|V|-1$ and $(|V| -1) (|V| - 2)/2$, respectively. Furthermore, $I_0(G)$ is MST of $G$ and $I_1(G)$ is the remaining non-MST edge weights.
\label{thm:partition}
\end{theorem}
Theorem \ref{thm:partition} is a non-trivial statement and used in the development of our proposed topological learning framework.

\subsection{Topological loss}
\label{sec:toploss}
 
Since network topology is completely characterized by 0D and 1D barcodes, the topological similarity between two networks can be measured using the difference between such barcodes. We will modify Wasserstein distance to measure the barcode difference for the graph filtration \citep{clough2019explicit,cohen2010lipschitz,hu2019topology,kolouri2019generalized,rabin2011wasserstein} as follows. 

Let $\Theta=(V^{\Theta},w^{\Theta})$  and $P=(V^P, w^P)$  be two given networks. For now, we will assume that the two networks have the same size, i.e., $|V^{\Theta}| = |V^P|$. The case $|V^{\Theta}| \neq |V^P|$ will be explained in Section \ref{sec:cardinality}. The 2-Wasserstein distance between 0D barcodes is given by
\begin{equation} 
\mathcal{L}_{0D} (\Theta, P) = \min_{\tau_0}  \sum_{\epsilon_{b} \in I_0(\Theta)} \big[ \epsilon_{b} - \tau_0(\epsilon_{b})\big]^2, 
\label{eq:L0D}
\end{equation}
where $\tau_0$ is a bijection from birth sets $I_0(\Theta)$ to $I_0(P)$. 
$\mathcal{L}_{0D} (\Theta, P)$ is termed {\em 0D topological loss}, which measures the topological dissimilarity between networks $\Theta$ and $P$ in terms of 0D features (connected components).
Recall that the death values of all connected components in networks are $\infty$. Thus, the 0D topological loss given in (\ref{eq:L0D}) simply ignores the death values. Even if we replace $\infty$ with a sufficiently large number, the loss will not be affected.

Similarly, the 2-Wasserstein distance between 1D barcodes is given by
\begin{equation} 
\mathcal{L}_{1D} (\Theta, P) =  \min_{\tau_1} \sum_{\epsilon_{d} \in I_1(\Theta)} \big[ \epsilon_{d} - \tau_1(\epsilon_{d})\big]^2,
\label{eq:L1D}
\end{equation}
where $\tau_1$ is a bijection from death sets $I_1(\Theta)$ to $I_1(P)$. 
$\mathcal{L}_{1D} (\Theta, P)$, termed {\em 1D topological loss}, measures the topological dissimilarity between networks $\Theta$ and $P$ in terms of cycles.
Again recall that the birth values of all cycles are $-\infty$. Thus, the 1D topological loss given in (\ref{eq:L1D}) simply ignores the birth values. Replacing $-\infty$ by a sufficiently small number will not affect the loss. 

Instead of trying to determine the  0D and 1D topological differences separately, we want a single summary statistic measuring the overall topological differences. This can be achieved by combining the 0D and 1D topological losses as
\[
\mathcal{L}_{top}(\Theta,P) = \mathcal{L}_{0D} (\Theta, P)  + \mathcal{L}_{1D} (\Theta, P).
\]
The 0D and 1D topological losses are variants of the {\em assignment problem}, which is typically solved using the cubic-time Hungarian algorithm \citep{edmonds1972theoretical}. However, for the graph filtration, the 0D and 1D losses have closed-form expressions that can be efficiently computed in $O \big( |I_0(\Theta)| \log |I_0(\Theta)| \big)$ and  $O \big( |I_1(\Theta)| \log |I_1(\Theta)| \big)$ time, respectively, as follows.

\begin{theorem}
\label{theorem:optimal}
For the 0D topological loss, we have
\[ \mathcal{L}_{0D} (\Theta, P) = \min_{\tau_0} \sum_{\epsilon_{b} \in I_0(\Theta)} \big[ \epsilon_{b} - \tau_0(\epsilon_{b})\big]^2 = \sum_{\epsilon_{b} \in I_0(\Theta)} \big[ \epsilon_{b} - \tau_0^*(\epsilon_{b})\big]^2,
\]
where $\tau_0^*$ maps the $i$-th smallest birth value in $I_0(\Theta)$ to the $i$-th smallest birth value in $I_0(P)$ for all $i$. Similarly for the 1D topological loss, we also have 
\[ \mathcal{L}_{1D} (\Theta, P) = \min_{\tau_1} \sum_{\epsilon_{d} \in I_1(\Theta)} \big[ \epsilon_{d} - \tau_1(\epsilon_{d})\big]^2 = \sum_{\epsilon_{d} \in I_1(\Theta)} \big[ \epsilon_{d} - \tau_1^*(\epsilon_{d}) \big]^2 ,\]
where $\tau_1^*$ maps the $i$-th smallest death value in $I_1(\Theta)$ to the $i$-th smallest death value in $I_1(P)$ for all $i$.
\label{thm:L1}
\end{theorem}

The minimizations in Theorem \ref{theorem:optimal} are equivalent to the following assignment problem. For monotonic sequences
$$a_1 < a_2  < \cdots < a_n, \quad b_1 < b_2  < \cdots < b_n,$$
we consider finding 
$\min_{\tau} \sum_{i=1}^n (a_i - \tau(a_i))^2$ over all possible bijections $\tau$. The optimal bijection is simply given by the identity permutation 
$\tau(a_i) = b_i$ and is proved by induction \citep{song.2021.MICCAI}. The problem statement can be extended to a more general assignment problem between different number of data
$$a_1 < a_2  < \cdots < a_m, \quad b_1 < b_2  < \cdots < b_n.$$
In this case, however, the analytical solutions given in Theorem \ref{theorem:optimal} are not applicable. The mismatching issue is discussed next.

\subsection{Topological loss between networks of different sizes}
\label{sec:cardinality}

Let $\Theta=(V^\Theta,w^\Theta)$ and $P=(V^P,w^P)$ be networks of different sizes, i.e., $|V^\Theta| \neq |V^P|$.  There is no bijection between the birth sets and between the death sets due to the mismatching issue. This problem is often addressed through data augmentation or empirical distribution methods \citep{bonneel2015sliced,carriere2017sliced,deshpande2018generative,karras2018progressive,kolouri2017optimal,liutkus2019sliced}. Data augmentation is probably the most popular technique for mismatches between topological features \citep{hu2019topology}, trees \citep{guo2020representations} and point sets \citep{chung2019exact,edelsbrunner2008persistent,ghrist2008barcodes,patrangenaru.2019,robins2016principal,xia2014persistent}. An alternative approach is to express the topological loss in terms of empirical distribution functions \citep{bonneel2015sliced,carriere2017sliced,deshpande2018generative,karras2018progressive,kolouri2017optimal,liutkus2019sliced}. 
The empirical distributions for birth sets of $\Theta$ and $P$ are given by
\begin{align*}
    \widehat F_\Theta(x) &= \frac{1}{|I_0(\Theta)|} \sum_{\epsilon_b \in I_0(\Theta)} \mathbb{1}_{\epsilon_b \leq x}, \\
    \widehat F_P(x) &= \frac{1}{|I_0(P)|} \sum_{\epsilon_b \in I_0(P)} \mathbb{1}_{\epsilon_b \leq x},
\end{align*}
where $\mathbb{1}_{\epsilon_b \leq x}$ is the indicator having the value 1 if $\epsilon_b \leq x$ and the value 0 otherwise. Their psedoinverses  $F_\Theta^{-1}(z)$ and $F_P^{-1}(z)$ are defined as 
the smallest $x$ for which $\widehat F_\Theta(x) \geq z$ and $\widehat F_P(x) \geq z$. Then 0D topological loss is given by \citep{kolouri2017optimal}
\begin{equation}
    \mathcal{L}_{0D}(\Theta,P) = \int_0^1\big(\widehat F_\Theta^{-1}(x) - \widehat F_P^{-1}(x)\big)^2dx.
    \label{eq:1dwass}
\end{equation}
Similarly, 1D topological loss can be defined in terms of the empirical distribution for death sets. We can efficiently compute the loss between networks of different sizes by computing the integral numerically \citep{song.2022}.
In addition, when the network sizes are equal, the empirical distributions are still well defined, and the loss given in (\ref{eq:1dwass}) exactly matches our analytical expression in Theorem \ref{theorem:optimal}.

In actual brain imaging applications, brain networks are usually constructed following standard pipeline resulting in networks with the same number of nodes \citep{fornito.2016,sporns.2003}. Thus, matching brain networks of different sizes can be avoided and there is no need to augment data or use empirical distributions. Comparing the two techniques, when the numerical accuracy relative to the definition of the Wasserstein distance is important, the data augmentation approach will not provide a relatively accurate distance. In such case, the  empirical distribution approach is preferable. Apart from the two major techniques for handling  networks of different sizes, there are  other available methods \citep{deshpande2018generative,karras2018progressive}. In particular, \citet{marchese2018signal} proposed a variant of the Wasserstein distance that explicitly penalizes cardinality and short-lived persistence points, and showed that the variant can utilize geometric information to improve the discriminative power in classification.

\subsection{Topological learning}

There are previous attempts to incorporate topology into learning and inference frameworks. \citet{adler.2017} proposed the parametric  model for persistence diagrams based on Gibbs distribution, and developed modeling, replication and inference procedures. \citet{marchese2018signal} built the probability measure on the metric space of persistence diagrams used in  classifying signals. In \citep{maroulas.2020}, Bayesian framework was developed by modeling persistence diagrams as a Poisson point process.
 \citet{naitzat.2020} investigated how Betti numbers change when different activation functions are used in deep neural networks.
 \citet{naitzat.2020} used deep neural networks as input to a topological learning framework. 
\citet{love.2021} proposed topological convolutional neural networks in deep learning. 

\begin{figure}[t]
\includegraphics[width=1\linewidth]{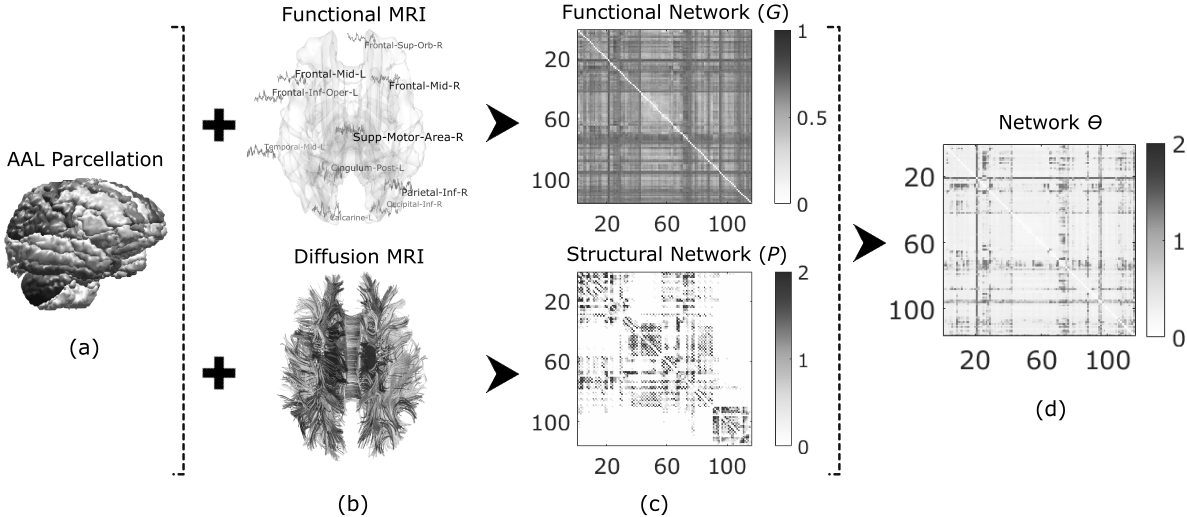}
\centering
\caption{Schematic illustrating topological learning for brain networks. (a) The Automated Anatomical Labeling (AAL) atlas obtained through structural-MRI is used to partition the human brain into 116 regions, which form nodes in brain networks. (b)-top In fMRI, brain activity at each node is measured as a time series of changes associated with the relative blood oxygenation level. (c)-top The functional connectivity between two nodes is given as the correlation between their fMRI time series, which goes through the metric transform, resulting in the functional network $G$. (b)-bottom The structural connectivity between two brain regions is measured by the number of white matter fiber tracts passing through them using dMRI. 
(c)-bottom Structural connectivities over all subjects are then normalized and scaled, resulting in the structural network $P$ that serves as the template where statistical analysis can be performed. The structural network $P$ is sparse while the functional network $G$ is densely connected. Since both networks are topologically different, it is difficult to integrate them together in a coherent model. Simply overlaying functional networks on top of the structural network will completely destroy 1D topology (cycles) of the functional networks. (d) The proposed framework learns network $\Theta$ that has the topological characteristics of both functional and structural networks.}
\label{fig:schematic_toplearning}
\end{figure}

In this paper, we incorporate topology into learning by minimizing 0D and 1D topological losses as follows. 
Let $G_1=(V, w^1),  \cdots ,G_n=(V, w^n)$ be observed networks used for training. Let $P=(V^P, w^P)$ be a network expressing a prior topological knowledge. In brain network analyses, $G_k$ can be a functional brain network of the $k$-th subject obtained from resting-state fMRI, and $P$ can be a template structural brain network obtained through dMRI. Functional brain networks are typically overlaid onto the template structural brain network (Figure \ref{fig:schematic_toplearning}) \citep{kang.2017,lv.2010,zhu.2014.NI}. Note that the node sets $V$ and $V^P$ may differ, which can happen in the situation where we try to integrate brain networks obtained from different parcellations.

We are interested in learning individual network model $\widehat \Theta_k$ from the $k$-th training network $G_k = (V, w^k)$ of the $k$-th subject by minimizing
\begin{equation}
\widehat \Theta_k =
\argmin_\Theta \mathcal{L}_F(\Theta,G_k) + \lambda_k \mathcal{L}_{top}(\Theta,P) ,
 \label{eq:training-sub}
\end{equation}
where squared Frobenius loss 
$\mathcal{L}_F(\Theta,G_k) = ||w^\Theta-w^k||^2_F$
is the goodness-of-fit term between the model and the individual observation, and parameter $\lambda_k$ controls the amount of topological information from network $P$ we are introducing into the model. The larger the value of $\lambda_k$, the more we learn toward topology of $P$. If $\lambda_k=0$, we no longer learn the topology of $P$ but fit the model only to the individual network $G_k$. Note that we learn toward the population average since $P$ is the template structural brain network.

\begin{figure}[t]
\includegraphics[width=1\linewidth]{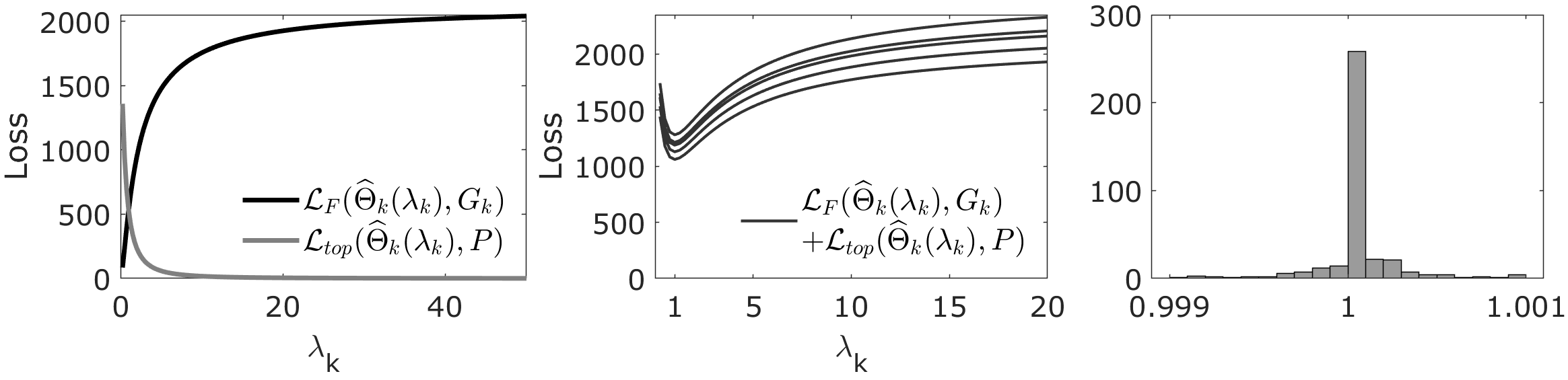}
\centering
\caption{Individual-level learning. Left: When $\lambda_k =0$, the learned network $\widehat \Theta_k$ is simply the individual network $G_k$.  As $\lambda_k$ increases, $\widehat \Theta_k$ is deformed such that the topology of $\widehat \Theta_k$ is closer to the structural template $P$. Middle: The sum of losses as a function  $\lambda_k$ for five representative subjects. Optimal $\lambda_k$ for each individual subject is the one that minimizes
the sum of losses. Right: Distribution of optimal $\lambda_k$ centered around $\lambda= 1.0000 \pm 0.0002$.}
\label{fig:loss_vs_lambda_plot}
\end{figure}

Determining optimal $\lambda_k$ for the $k$-th subject can be done as follows.
For each pre-specified $\lambda_k$, we find optimal $\widehat \Theta_k$ by minimizing the objective function (\ref{eq:training-sub}) over all possible $\Theta$. We then determine which $\lambda_k$ and its corresponding $\widehat \Theta_k$ give the minimum sum of losses $\mathcal{L}_F+\mathcal{L}_{top}$. Figure \ref{fig:loss_vs_lambda_plot}-left displays the losses $\mathcal{L}_F$ and $\mathcal{L}_{top}$ as a function of $\lambda_k$. Figure \ref{fig:loss_vs_lambda_plot}-middle displays the sum of losses as a function of $\lambda_k$ for five representative subjects. 
Figure \ref{fig:loss_vs_lambda_plot}-right displays the histogram of the distribution of optimal $\lambda_k$ based on all subjects. 
The average optimal $\lambda_k$ over all subjects is $\lambda_k= 1.0000 \pm 0.0002$ showing highly stable result. Such stable result is not plausible for non-topological loss functions. Similar to the well-known stability result in the persistent homology literature \citep{cohen2010lipschitz}, we can algebraically show
$$\mathcal{L}_{top}(\Theta,P) \leq C ||w^\Theta-w^P||_F^2 $$ for some $C$ providing the stability of topological loss. 
For real data used in this paper, we have the least upper bound of $0.4102$ for all subjects.

\begin{figure}[t]
\includegraphics[width=1\linewidth]{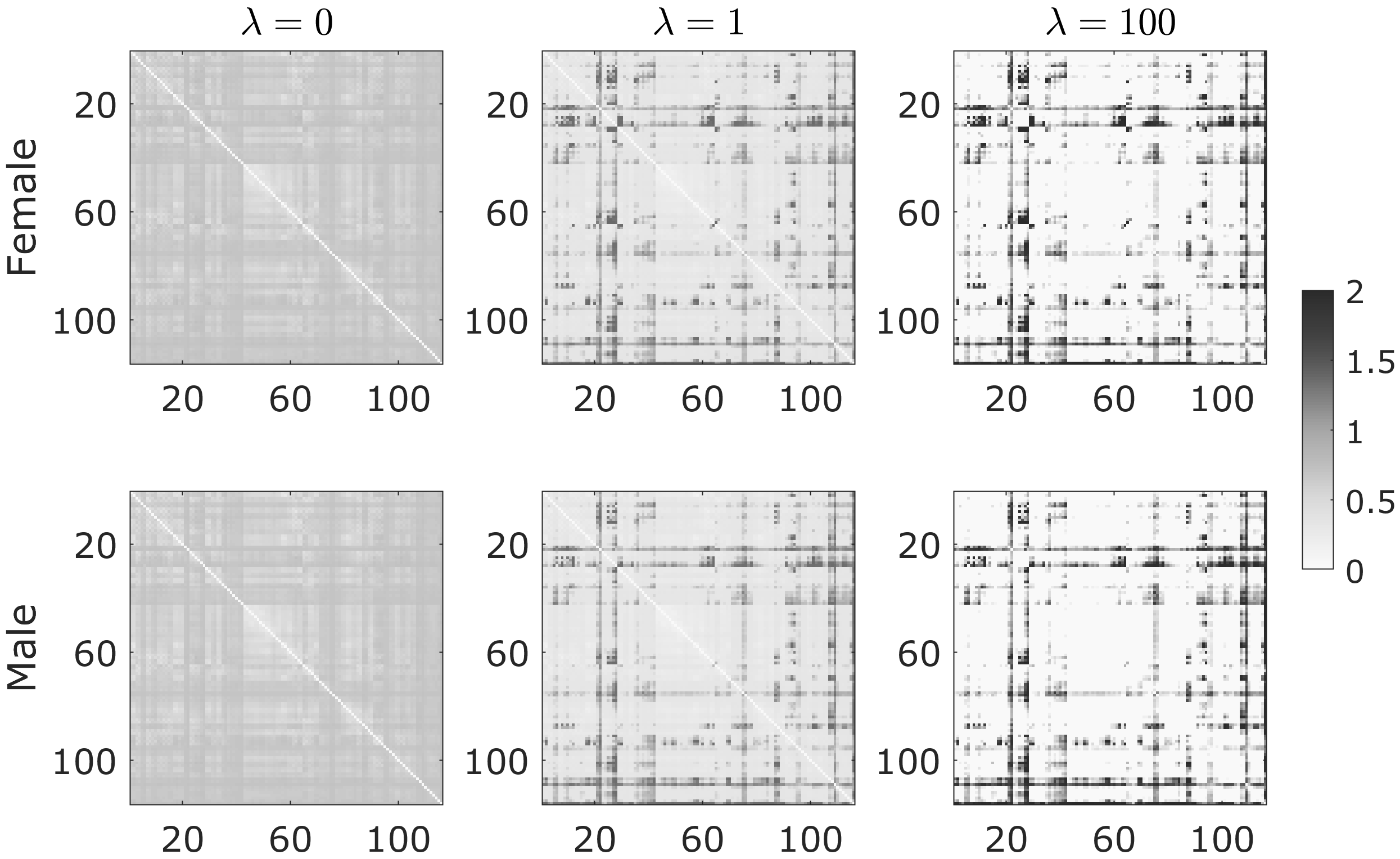}
\centering
\caption{Group-level networks of female (top row) and male (bottom row) are estimated by minimizing the objective function (\ref{eq:training-group}) with different $\lambda=0,1$ and 100.}
\label{fig:group_network}
\end{figure}

\begin{figure}[t]
     \centering
     \begin{subfigure}[b]{\textwidth}
         \centering
         \includegraphics[width=1\textwidth]{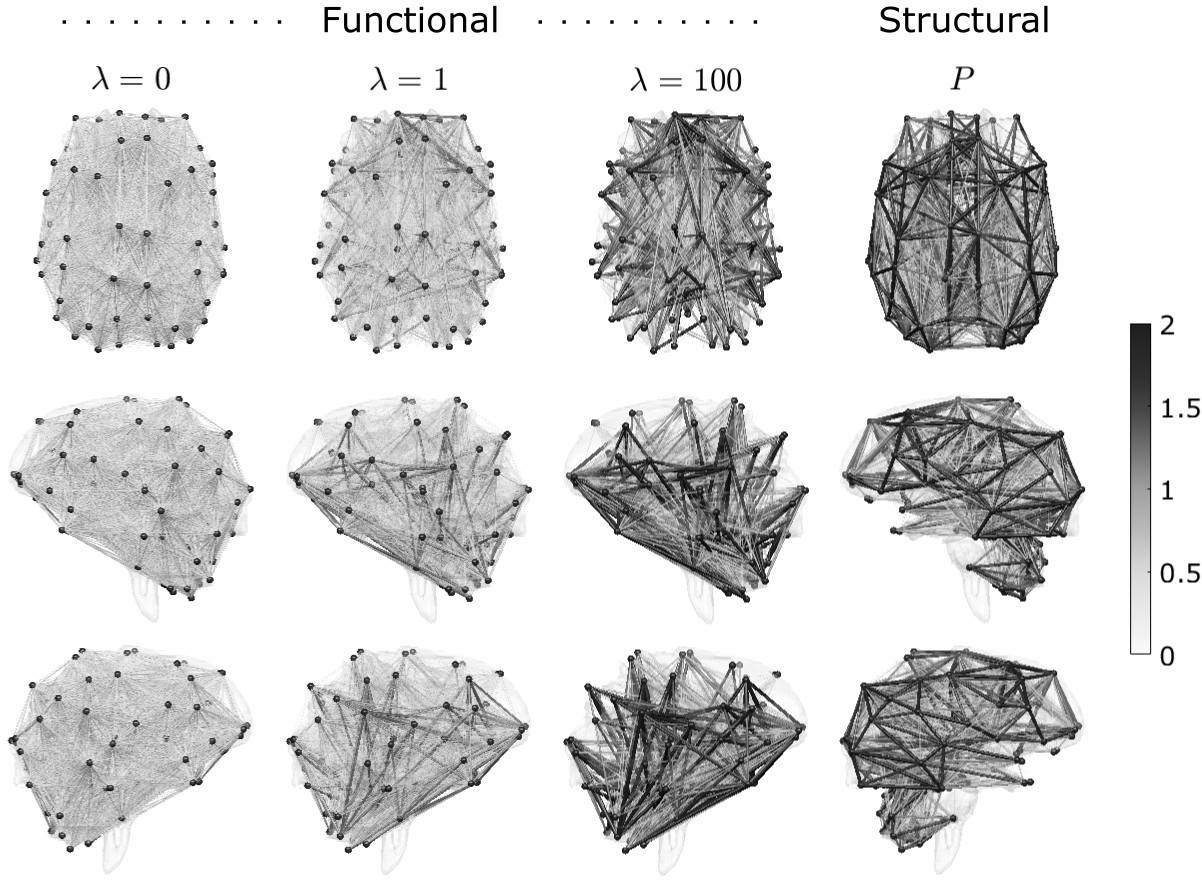}
         \caption{}
         \label{fig:cycles_a}
     \end{subfigure}
     \par\bigskip
     \begin{subfigure}[b]{\textwidth}
         \centering
         \includegraphics[width=1\textwidth]{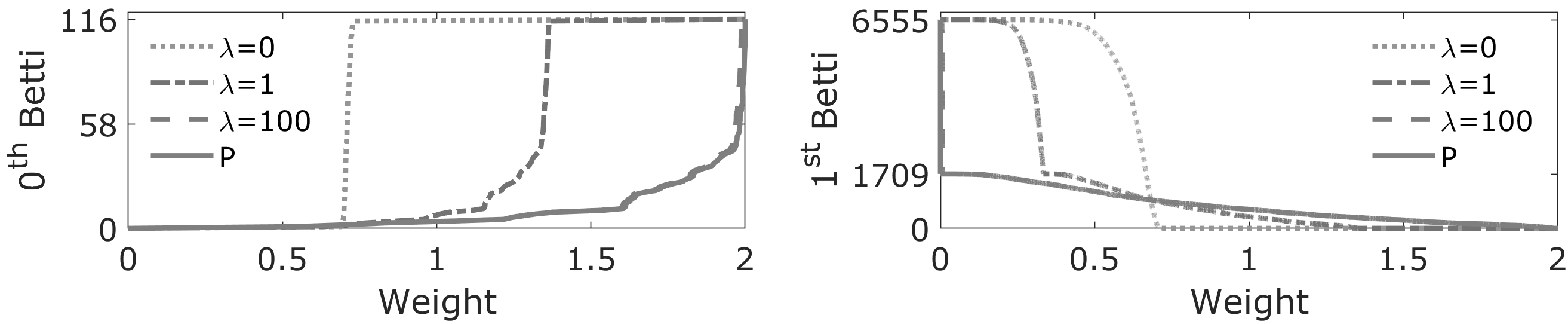}
         \caption{}
         \label{fig:cycles_b}
     \end{subfigure}
    \caption{(a) Group-level networks learned by minimizing the objective function (\ref{eq:training-group}) over all subjects with $\lambda=0,1$ and 100. Fixed $\lambda$ is chosen across all the subjects. The template structural network $P$ is shown in the last column.
    (b) As $\lambda$ increases, Betti-plots of the group-level network are adjusted toward that of $P$. $\beta_0$-plot shows that the connected components in the structural network $P$ are gradually born over a wide range of edge weights during the graph filtration. $\beta_1$-plot shows topological sparsity of lack of cycles in the structural network $P$. While the group-level functional network (when $\lambda=0$) is densely connected with up to 6555 number of cycles, the structural network is sparsely connected with only 1709 cycles. }
    \label{fig:cycles}
\end{figure}

Although group-level learning is not the focus in this paper, we can also learn a model $\widehat \Theta$ using multiple training networks such that
\begin{equation}
    \widehat \Theta =
\argmin_\Theta \frac{1}{n}\sum_{k=1}^n \mathcal{L}_F(\Theta,G_k) + \lambda \mathcal{L}_{top}(\Theta,P), 
 \label{eq:training-group}
\end{equation}
where single optimal parameter $\lambda$ is chosen. Figure \ref{fig:group_network} displays average networks of females and males while Figure \ref{fig:cycles} displays average networks over all subjects by minimizing the objective function (\ref{eq:training-group}) with $\lambda=0, 1$ and 100. The larger the value of $\lambda$, the more we reinforce the average networks with topology of the template structural brain network. Even though we will not show in this paper, statistical significance of network differences between females and males can be determined by using the exact topological inference developed in our previous work \citep{chung2017exact,chung2019exact}.

\subsection{Averaging networks of different sizes and topology}

Another application of the proposed topological loss is learning the average of networks with different sizes and topology, which is a difficult task using existing methods.
Given $n$ networks  $G_1=(V_1, w^1), \cdots,G_n=(V_n, w^n)$ with different node sets, we are interested in obtaining its {\em topological mean}. Since the size and topology of the networks are different, we cannot directly average edge weight matrices $w^1, \cdots, w^n$. 
Motivated by the Fr\'{e}chet mean  \citep{le.2000,turner.2014,zemel.2019}, here we obtain the topological mean $\widehat \Theta$ by minimizing the sum of 0D and 1D topological losses
\[
\widehat \Theta =  \argmin_{\Theta} \sum_{k=1}^n \mathcal{L}_{top}(\Theta, G_k)  =   \argmin_{\Theta} \sum_{k=1}^n   \Big[ \mathcal{L}_{0D}(\Theta, G_k) + \mathcal{L}_{1D}(\Theta, G_k) \Big],
\]
where $\widehat \Theta$ is a network viewed as the topological centroid of given $n$ networks. The optimization can be done analytically as follows \citep{rabin2011wasserstein}. 

The 0D topological loss  $\mathcal{L}_{0D}$ depends on the birth values of networks $G_1, \cdots, G_n$. 
For networks with the same number of nodes, we have the same $m_0$ number of birth values.
Let  $b_{k1} < b_{k2} < \cdots < b_{k m_0}$ be the birth values of network $G_k$. Let  $\theta_1 < \theta_2 < \cdots < \theta_{m_0}$ be the birth values of network $\Theta$. By Theorem \ref{theorem:optimal}, the sum of 0D losses is  equivalent to
\[
\sum_{k=1}^n \mathcal{L}_{0D}(\Theta, G_k)  =  \sum_{k=1}^{n}  (\theta_1 - b_{k1})^2 +  \sum_{k=1}^{n}  (\theta_2 - b_{k2})^2 +  \cdots +  \sum_{k=1}^{n}  (\theta_{m_0} - b_{ km_0})^2,
\]
which is quadratic so we can find the minimum by setting its derivative equal to zero. The solution is given by  
$ \widehat \theta_j = \sum_{k=1}^n b_{kj} \big/ n.$
Similarly, the sum of 1D losses is equal to the sum of squared differences of death values.
Thus, the $i$-th smallest birth (or death) value of the topological mean network $\widehat \Theta$ is equal to the mean of all the $i$-th smallest birth (or death) values of the given $n$ networks. For networks with different number of nodes, we can match networks through the empirical distribution method described in Section \ref{sec:cardinality}.

\begin{figure}[t]
\includegraphics[width=1\linewidth]{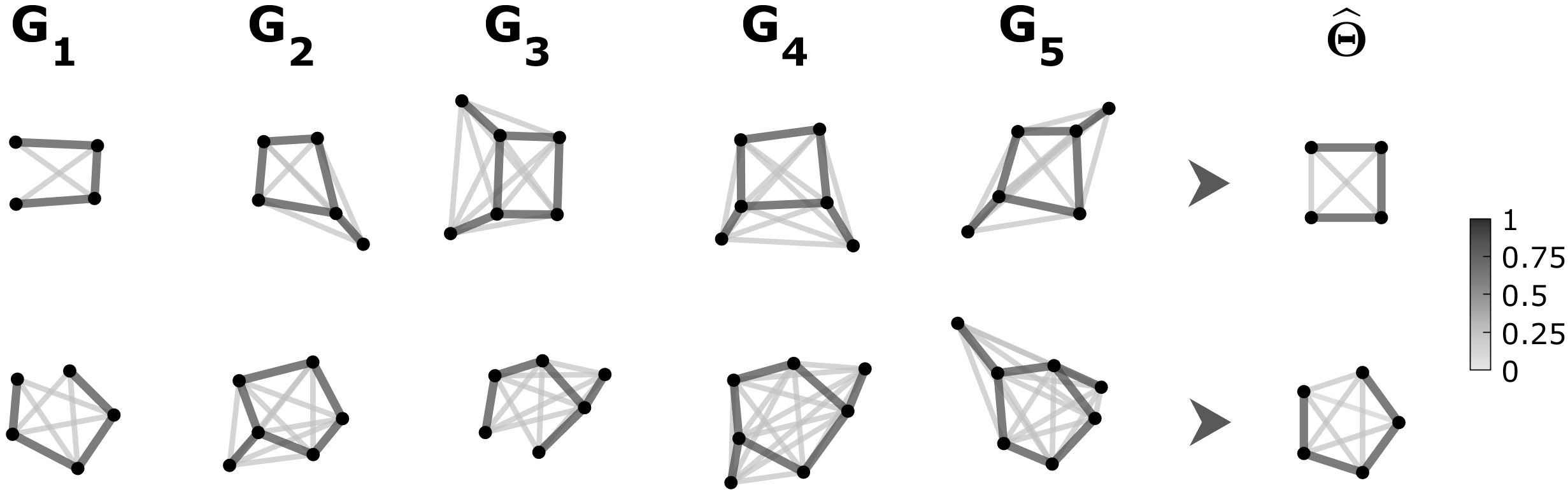}
\centering
\caption{Examples of averaging networks of different sizes and topology using the empirical distribution method in Section \ref{sec:cardinality}. The topological mean network $\widehat\Theta$ (right) is the topological centroid of five networks $G_1, \cdots, G_5$, showing the average topological pattern. The topological mean network $\widehat \Theta$ is estimated by minimizing the sum of topological losses $\min_{\Theta} \sum_{k=1}^5 \mathcal{L}_{top}(\Theta, G_k)$. The topological mean network $\widehat \Theta$ highlights topological characterization of the five networks. The existing methods will have difficulty averaging networks of different sizes and topology.}
\label{fig:meantop}
\end{figure}

Given all the birth and death values of the topological mean network, we can completely recover its topology. However, the network is only unique in the topological sense but not in a geometric sense. We can have multiple different networks that are geometrically different but with identical topology. For two networks $A$ and  $B$ whose edge weights are different, we can have identical birth sets $I_0(A) = I_0(B)$ and identical death sets $I_1(A) = I_1(B)$ so that $\mathcal{L}_{top}(A,C) = \mathcal{L}_{top}(B,C)$ for some network $C$. Such examples can be obtained by permuting node labels or rotating networks geometrically. However, the non-uniqueness may not be disadvantageous in classification and image segmentation. By rotating graphs embedded in 2D images, the number of training samples can be drastically increased, boosting the classification and segmentation performance \citep{marcos.2016,taylor.2018}. For brain network application studied in this paper, we are introducing the Frobenius loss that constrains the networks geometrically to avoid this geometric ambiguity. Figure \ref{fig:meantop} illustrates toy examples of averaging networks of different sizes and topology. Since we can have many differently shaped networks that are topologically equivalent, it is not possible to identify $\widehat \Theta$ uniquely with the averaged birth and death values. For Figure \ref{fig:meantop}-top example, it would be possible for the topological mean network $\widehat\Theta$ to have the three darker edges extending from one node to the other three remaining nodes.

\subsection{Numerical implementation}

The topological learning given in (\ref{eq:training-sub}) estimates $\Theta = (V^{\Theta}, w^{\Theta})$ iteratively through gradient descent \citep{bottou1998online}. The gradient of the topological loss can be computed efficiently without taking numerical derivatives. In particular, computation of the gradient mainly comprises computing barcodes $I_0$ and $I_1$, and finding the optimal matchings through Theorem \ref{theorem:optimal}. The gradient of the topological loss  $\nabla\mathcal{L}_{top}(\Theta,P)$ with respect to edge weights $w^{\Theta} = (w_{ij}^{\Theta})$ is given as a gradient matrix whose $ij$-th entry is 
\begin{equation*}
\frac{\partial \mathcal{L}_{top}(\Theta,P)}{\partial w_{ij}^{\Theta}} 
= 
\begin{cases}
2 \big[ w_{ij}^{\Theta} - \tau_0^*( w_{ij}^{\Theta}  ) \big] & \text{if } w_{ij}^{\Theta} \in I_0(\Theta); \\
2 \big[ w_{ij}^{\Theta}  - \tau_1^*(w_{ij}^{\Theta} ) \big] & \text{if } w_{ij}^{\Theta} \in I_1(\Theta),
\end{cases}
\end{equation*}
since $I_0(\Theta)$ and $I_1(\Theta)$ partition the weight set (Theorem \ref{thm:partition}). By slightly adjusting the edge weight $w_{ij}^{\Theta}$, we have the slight adjustment of either a birth value in 0D barcode or a death value in 1D barcode, which in turn alters the topology of the network. 
During the estimation of $\Theta$, we take steps in the direction of  negative gradient:
\[
w^{\Theta}_{ij} \leftarrow w^{\Theta}_{ij} - 0.1 \Big( 2(w^{\Theta}_{ij}-w^k_{ij}) + \lambda \frac{\partial \mathcal{L}_{top}(\Theta,P)}{\partial w_{ij}^{\Theta}} \Big),
\]
where 0.1 is the learning rate. As $w_{ij}^{\Theta}$ moves closer to its optimal match, the topology of the estimated network $\Theta$ gets closer to that of $P$ while the Frobenius norm keeps the estimation $\Theta$ close to the observed network $G_k$.

Finding 0D birth values $I_0(G)$ is {\em equivalent} to finding edge weights comprising the {\em maximum spanning tree} (MST) of $G$ \citep{lee2012persistent}. Once $I_0$ is computed, $I_1$ is simply given as the rest of the edge weights (Theorem \ref{thm:partition}). Then, we can compute the optimal matchings $\tau_0^*$ and $\tau_1^*$ between $\Theta$ and $P$ by matching edge weights in the ascending order. The computational complexity of the topological loss gradient is dominated by the computation of the MST using standard algorithms such as Prim's and Kruskal's, which take $O( |E| \log|V|)$ run-time with $|E|$ number of edges and $|V|$ number of vertices.

Many efficient algorithms for Wasserstein distance computation often utilize the geometric structure of data \citep{sharathkumar.2012}.  In particular, \citet{kerber2017geometry}  proposed an approximation algorithm based on $k$-d trees with the time complexity empirically estimated as $O(n^{1.6})$ for $n$ scatter points in an arbitrary persistence diagram. The estimation is obtained using linear regression on observed running time as a function of $n$.
Translated into the graph filtration setting with $|E|$ number of edges, this is equivalent to $O(|E|^{1.6})$. 
The approximation algorithm in \citep{kerber2017geometry} is much faster than the Hungarian algorithm with $O(|E|^3)$, but still slower than our exact Wasserstein computation method with $O( |E| \log|V|)$. Note there are $|E| = |V| (|V|-1)/2$ edges. Our algorithm exploits the geometric structure of the graph filtration, resulting in the persistence diagram representation in the form of 1-dimensional sorted scalar values. This exploitation enables us to compute the Wasserstein distances exactly at the faster run time.

\section{Validation}

We performed two simulation studies to assess the performance of the topological loss as a dissimilarity measure between networks of different topology. Networks of the same size were simulated since many existing methods cannot handle networks of different sizes.

\subsection{Study 1: Random network model with ground truth}

Initial data vector $b_i$ at node $i$ was simulated as independent and identically distributed multivariate normal across $n$ subjects, i.e., $b_i \sim \mathcal{N}(0,I_n)$ with the identity matrix $I_n$ as the covariance matrix of size $n \times n$. The new data vector $x_i$ at node $i$ was then generated by introducing additional dependency structures to $b_i$ through a mixed-effects model that partitions the covariance matrix of $x_i$ into $c$ blocks forming modular structures \citep{chung2019exact,snijders1995use}:
\begin{align*}
x_1,...,x_a &= b_1 + \mathcal{N}(0,\sigma^2I_n), \\
x_{a+1},...,x_{2a} &= b_{a+1} + \mathcal{N}(0,\sigma^2I_n), \\
&\vdots \\
x_{(c-1)a+1},...,x_{ca} &= b_{(c-1)a+1} + \mathcal{N}(0,\sigma^2I_n),
\end{align*}
where $a$ is the number of nodes in each module.
 
\begin{figure}[t]
\includegraphics[width=1\linewidth]{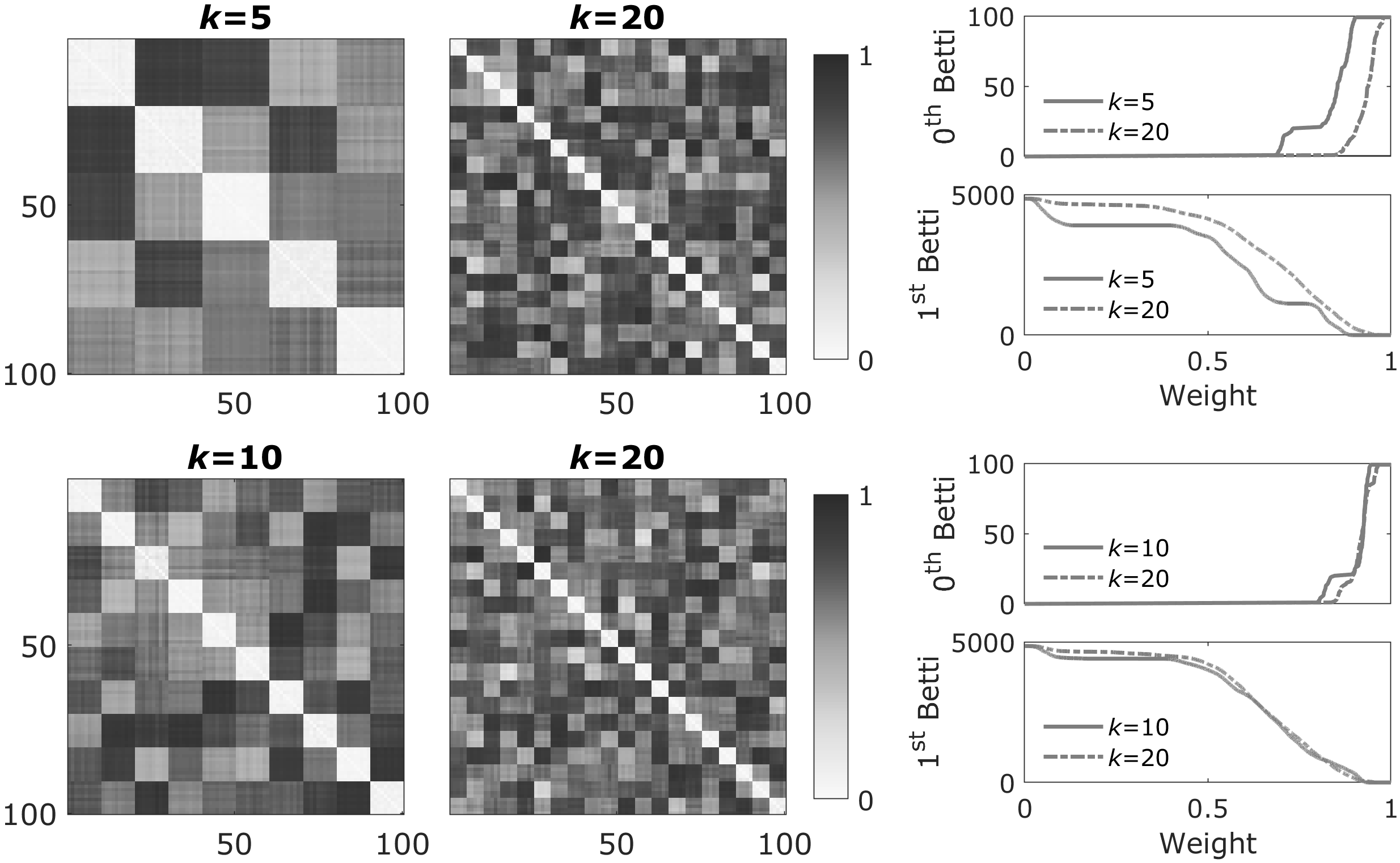}
\centering
\caption{Simulation study 1 with small noise ($\sigma=0.1$). The comparison of networks with different topology: $k=5$  vs. 20 (first row) and 10 vs. 20 (second row). The change of Betti numbers over filtration values shows topological differences. The topological difference in the second row is more subtle compared to the first row.}
\label{fig:sim1}
\end{figure}

In this simulation, networks with 100 nodes are partitioned into $c$ modules, where $c$ is chosen such that 100 is divisible by $c$. This choice of $c$ makes the partitioning of each network into modules straightforward. In particular,  $c= 2, 5, 10, 20$ number of modules are chosen, and thus there are $a = 50, 20, 10, 5$ nodes in each module, respectively. The simulation is done in small noise ($\sigma=0.1$) and large noise ($\sigma=0.5,1$) settings. Note that it gets more difficult to discriminate between different networks as the variability increases. We compute the Pearson correlation coefficient $\rho^x_{ij}$ between $x_i$ and $x_j$, which is then translated and scaled as metric $w_{ij}^x=\sqrt{(1-\rho^x_{ij})\big/2}$ \citep{chung2019exact}. This gives a modular network $\mathcal{X}=(V,w^x)$.
Figure \ref{fig:sim1} shows examples of simulated modular networks.

Based on the statistical model above, we simulated two groups of networks consisting of $n=7$ subjects in each group. The small sample size was chosen such that exact permutation tests can be done by generating exactly ${14 \choose 7}=3432$ number of every possible permutation. 
We employed the exact permutation test for network distances \citep{chung2019exact} on the two-sample $t$-statistic to evaluate the performance of the topological loss $\mathcal{L}_{top}$ relative to several baseline measures.
The separate evaluation of $\mathcal{L}_{0D}$ and $\mathcal{L}_{1D}$ helps with understanding the relative performance of the individual 0D and 1D Wasserstein distances in comparison with $\mathcal{L}_{top}$.  Thus, we additionally evaluated the separate performance of $\mathcal{L}_{0D}$ and $\mathcal{L}_{1D}$. For comparison, we tested the topological loss against Euclidean losses such as $\mathcal{L}_1$-, $\mathcal{L}_2$- and $\mathcal{L}_\infty$-norms. We further compared against other topological distances such as bottleneck, Gromov-Hausdorff (GH) and Kolmogorov-Smirnov (KS) distances \citep{chazal2009gromov,chung2019exact,cohen2007stability}. The bottleneck and GH-distance are widely-used baseline distances in persistent homology \citep{carlsson2010characterization} and brain networks \citep{lee2012persistent}. KS-distance based on $\beta_0$ and $\beta_1$ curves is later introduced as an interpretable alternative \citep{chung2017topological}.

\paragraph*{Network difference} We compared networks with different number of modules 2 vs. 10, 5 vs. 20 and 10 vs. 20. Since the networks had different topological structures, the distances were expected to detect the differences (Figure \ref{fig:sim1}). The simulations were independently performed 50 times and the performance results were given in terms of the false negative rate computed as the fraction of 50 simulations that gave $p$-values above 0.05 (Table \ref{tab:sim1}). In general, the topological loss performed relatively well for different noise settings. When the topological difference is obvious ($\sigma=0.1$), the proposed method performed exceptionally well. However, in the large noise settings ($\sigma=0.5,1$), all the distances including the topological loss did not perform well except the KS-distance. Since KS-distances measures the maximum difference over both Betti numbers and filtration values, it is sensitive to topological differences in networks. The sensitivity of KS-distance is advantageous in the high noise settings. On the other hand, $\mathcal{L}_{top}$ mainly penalizes differences in filtration values where topological changes occur, and thus is less sensitive than the KS-distance. In addition, $\mathcal{L}_{1D}$ performed substantially better than $\mathcal{L}_{0D}$ in all noise settings. The higher discriminative power of $\mathcal{L}_{1D}$ is expected since there are many more cycles than connected components in networks. In comparison, $\mathcal{L}_{top} = \mathcal{L}_{0D} + \mathcal{L}_{1D}$ demonstrated the increase in performance in high noise settings ($\sigma=0.5,1$). The improved performance suggests that the combined use of $\mathcal{L}_{0D}$ and $\mathcal{L}_{1D}$ is complementary, and thus $\mathcal{L}_{top}$ is the better choice for differentiating overall network topology.

\paragraph*{No network difference} We compared networks with  the number of modules  2 vs. 2, 5 vs. 5 and 10 vs. 10, which should result in networks of similar topology in each group. It was expected that the networks were not topologically different and we should not detect any signal. The simulations were independently performed 50 times and the performance results were given in terms of the false positive rates computed as the fraction of 50 simulations that gave $p$-values below 0.05 (Table \ref{tab:sim1}). While most of the baseline distances performed relatively well when there was no network difference, they had the tendency to produce false negatives. The KS-distance did not perform well in all cases and produced false positives even in the low noise setting ($\sigma=0.1$). The overly sensitive nature of the KS-distance was responsible for the huge false positives. On the other hand, $\mathcal{L}_{top}$ was significantly more robust than the KS-distance with much less false negatives, even in the high noise setting ($\sigma=1$). There were no significant differences in the performance among $\mathcal{L}_{0D}, \mathcal{L}_{1D}$ and $\mathcal{L}_{top}$. The almost identical performance suggests that $\mathcal{L}_{top}$ inherits robustness from $\mathcal{L}_{0D}$ and $\mathcal{L}_{1D}$, and is expected to perform relatively well.

\begin{table}[t]
\caption{Study 1. The performance results are summarized  in terms of false negative rates (2 vs. 10, 5 vs. 20, 10 vs. 20) and false positive rates (2 vs. 2, 5 vs. 5, 10 vs. 10). Smaller numbers are better.}
\label{tab:sim1}
\centering
\begin{tabular}{cc|cccccccccc}
\toprule
$\sigma$ & $c$ & $\mathcal{L}_1$ & $\mathcal{L}_2$ & $\mathcal{L}_\infty$ & GH & Bottleneck & KS($\beta_0$) & KS($\beta_1$) & $\mathcal{L}_{0D}$ & $\mathcal{L}_{1D}$ & $\mathcal{L}_{top}$ \\ 
\midrule
\multirow{6}{*}{0.1}
& 2 vs. 10 & 0.02 & 0.00 & 0.02 & 0.30 & 0.40 & 0.00 & 0.00  & 0.70 & 0.02 & 0.02 \\
& 5 vs. 20 & 0.06 & 0.00 & 0.02 & 0.20 & 0.32 & 0.02 & 0.00  & 0.48 & 0.00 & 0.00 \\
& 10 vs. 20 & 1.00 & 0.86 & 0.34 & 0.32 & 0.20 & 0.24 & 0.00 & 0.78 & 0.08 & 0.08 \\ \cmidrule{3-12}
& 2 vs. 2 & 0.00 & 0.00 & 0.00 & 0.00 & 0.00 & 0.54 & 0.88   & 0.00 & 0.00 & 0.00 \\
& 5 vs. 5 & 0.00 & 0.00 & 0.00 & 0.00 & 0.00 & 0.10 & 0.42   & 0.00 & 0.00 & 0.00 \\
& 10 vs. 10 & 0.00 & 0.00 & 0.00 & 0.00 & 0.00 & 0.02 & 0.12 & 0.00 & 0.00 & 0.00 \\ 

\bottomrule \toprule

\multirow{6}{*}{0.5}
& 2 vs. 10 & 0.04 & 0.02 & 0.42 & 0.84 & 0.60 & 0.00 & 0.00  & 0.66 & 0.12 & 0.10 \\
& 5 vs. 20 & 0.16 & 0.08 & 0.58 & 0.82 & 0.72 & 0.02 & 0.00  & 0.70 & 0.10 & 0.08 \\
& 10 vs. 20 & 1.00 & 0.98 & 0.80 & 0.92 & 0.82 & 0.06 & 0.00 & 0.88 & 0.68 & 0.62 \\ \cmidrule{3-12}
& 2 vs. 2 & 0.00 & 0.00 & 0.00 & 0.02 & 0.00 & 0.80 & 0.56   & 0.00 & 0.00 & 0.00 \\
& 5 vs. 5 & 0.00 & 0.00 & 0.00 & 0.02 & 0.00 & 0.92 & 0.58   & 0.00 & 0.00 & 0.00 \\
& 10 vs. 10 & 0.00 & 0.00 & 0.00 & 0.02 & 0.00 & 0.98 & 0.84 & 0.00 & 0.00 & 0.00 \\ 

\bottomrule \toprule

\multirow{6}{*}{1}
& 2 vs. 10 & 0.12 & 0.14 & 0.92 & 0.90 & 0.68 & 0.02 & 0.00  & 0.72 & 0.26 & 0.24 \\
& 5 vs. 20 & 0.74 & 0.62 & 0.92 & 0.94 & 0.84 & 0.06 & 0.04  & 0.82 & 0.58 & 0.56 \\
& 10 vs. 20 & 1.00 & 1.00 & 1.00 & 0.94 & 0.96 & 0.06 & 0.38 & 0.92 & 0.90 & 0.86 \\ \cmidrule{3-12}
& 2 vs. 2 & 0.00 & 0.00 & 0.02 & 0.02 & 0.08 & 0.88 & 0.40   & 0.02 & 0.00 & 0.00 \\
& 5 vs. 5 & 0.00 & 0.00 & 0.00 & 0.06 & 0.00 & 0.92 & 0.52   & 0.00 & 0.00 & 0.00 \\
& 10 vs. 10 & 0.00 & 0.00 & 0.02 & 0.06 & 0.00 & 0.92 & 0.74 & 0.00 & 0.00 & 0.00 \\ 
\bottomrule
\end{tabular}
\end{table}

\subsection{Study 2: Comparison against graph matching algorithms}

The aim of this simulation is to evaluate the performance of the proposed topological matching process against existing graph matching algorithms \citep{cho2010reweighted,gold1996graduated,leordeanu2005spectral,leordeanu2009integer,zhou2013deformable}. Given weighted networks $G_1=(V_1,w^1)$ and $G_2=(V_2,w^2)$, we need to find mapping $\tau_{gm}$ between nodes $i_1,j_1\in V_1$ and $i_2,j_2\in V_2$ that best preserves edge attributes between edge weights $w^1_{i_1j_1}\in w^1$ and $w^2_{i_2j_2} \in w^2$. We seek $\tau_{gm}$ to  maximize the graph matching cost
\[ J(\tau_{gm}) = \sum_{w^1_{i_1j_1},w^2_{i_2j_2}} f(w^1_{i_1j_1},\tau_{gm}(w^2_{i_2j_2})),  \]
where $f$ measures the similarity between edge attributes, and the summation is taken over all possible edge weights. 
The matching cost $J(\tau_{gm})$ quantifies similarity between networks by taking large values for similar networks and values close to zero for dissimilar networks; hence, $J(\tau_{gm})$ is somewhat the inverse of distance metrics.
We compared the proposed topological loss against four graph matching algorithms: graduated assignment (GA) \citep{gold1996graduated}, spectral matching (SM) \citep{leordeanu2005spectral}, integer projected fixed point method (IPFP) \citep{leordeanu2009integer} and re-weighted random walk matching (RRWM) \citep{cho2010reweighted}. 
Such graph matching methods are widely used  in medical imaging, computer vision and machine learning studies \citep{cour2006balanced,tian2012convergence,wang2020learning,yu2018generalizing,zhang2019kergm,zhou2013deformable}. For all the baseline methods, we used existing implementations from authors' repository websites listed in the publication. We also used parameters recommended in the public code for each baseline algorithm {\em without} modification. Since we are dealing with weighted edges, the graph matching algorithms based on binary edge weight are excluded in the study \citep{babai.1983, guo2020representations,zavlanos.2008}.

\begin{figure}[t]
\includegraphics[width=.9\linewidth]{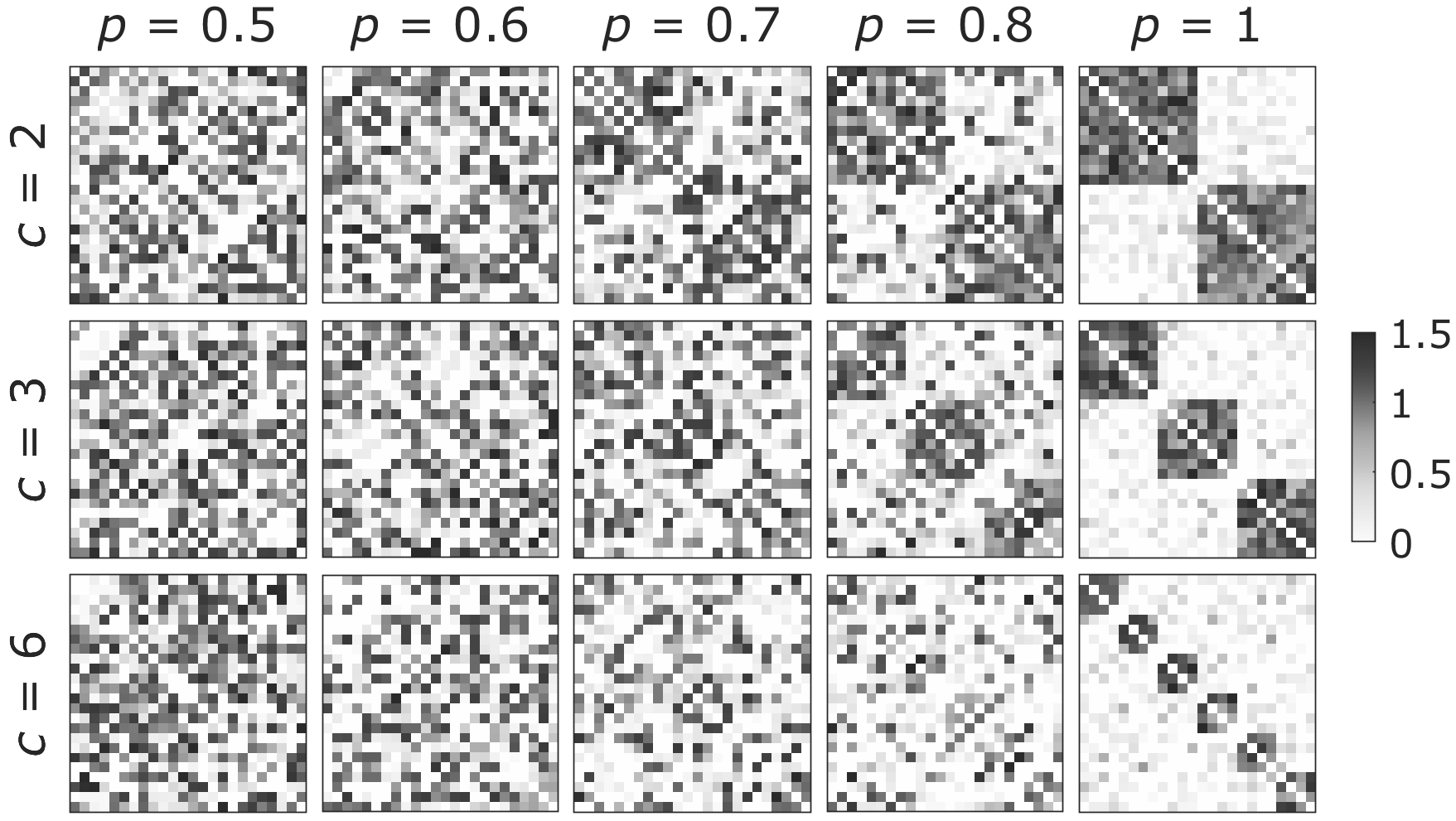}
\centering
\caption{Study 2 simulation examples. Network modular structures vary as parameter $p$ (probability of connection within modules) and parameter $c$ (the number of modules) change. The modular structure becomes more pronounced as $p$ increases.}
\label{fig:modular}
\end{figure}

\begin{figure}[t]
\includegraphics[width=.7\linewidth]{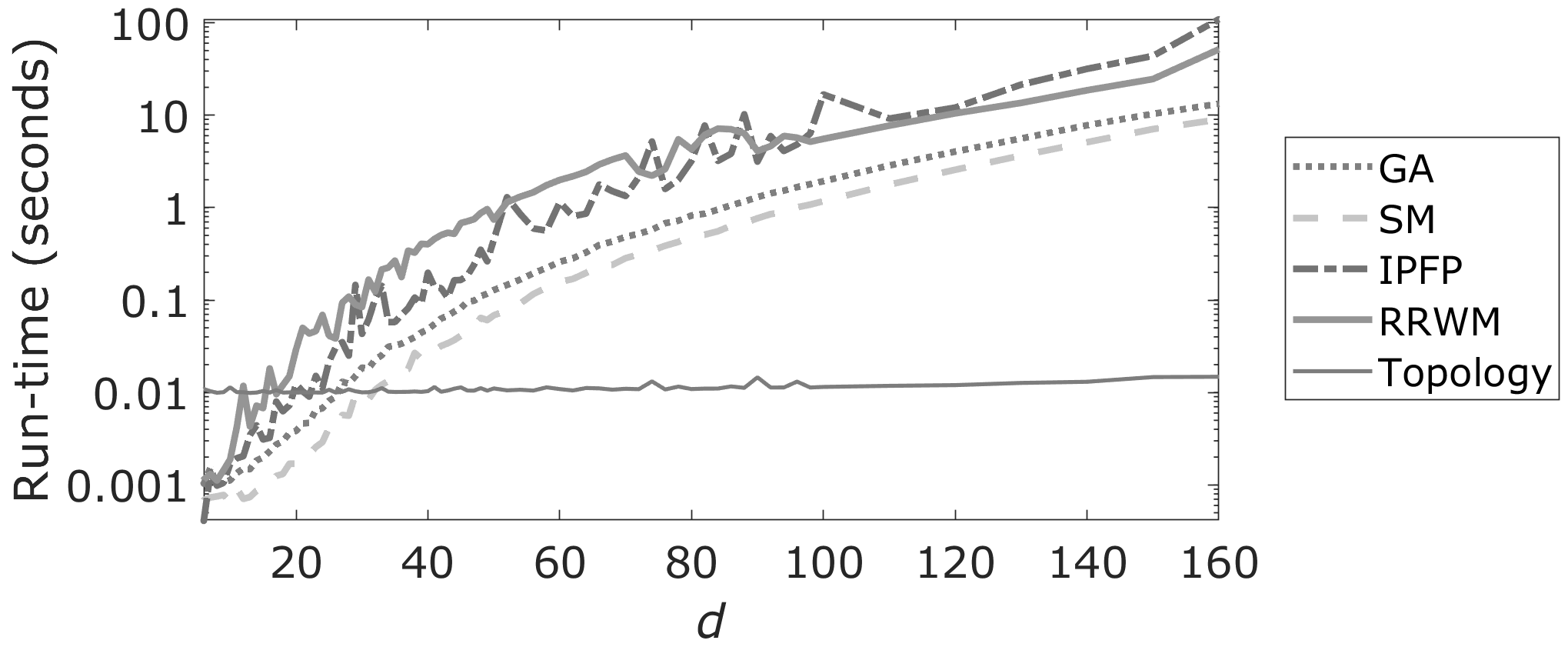}
\centering
\caption{Study 2 run-time given in the logarithmic scale.  The average runtime measures the amount of time each algorithm takes to compute its matching cost between two modular networks of size $d$ starting from edge weights as a given input. The run-time performance of the baseline methods is consistent with \citet{cour2006balanced} for GA and SM, and \citet{zhang2019kergm} for IPFP and RRWM.}
\label{fig:runtime}
\end{figure}

In study 2, a different random network model from study 1 is used. We simulated a random modular network $\mathcal{X}$ with $d$ number of nodes and $c$ number of modules, where the nodes are evenly distributed among modules. Figure \ref{fig:modular} displays modular networks with $d=24$ nodes and $c=2,3,6$ modules such that we have $d/c=12,8,4$ number of nodes in each module, respectively.
Since the time complexity of the aforementioned graph matching baselines can be very demanding (Figure \ref{fig:runtime}), we only considered $d=12,18,24$ and $c=2,3,6$ in this simulation. 
Each edge connecting two nodes within the same module was then assigned a random weight following a normal distribution $\mathcal{N}(\mu,\sigma^2)$ with probability $p$, and Gaussian noise $\mathcal{N}(0,\sigma^2)$ with probability $1-p$.
Edge weights connecting nodes between different modules had probability $1-p$ of being $\mathcal{N}(\mu,\sigma^2)$, and probability $p$ of being $\mathcal{N}(0,\sigma^2)$. With a larger value of within-module probability $p$, we have a more pronounced modular structure. Any negative edge weights were set to zero. This gives the random network $\mathcal{X}$ that exhibits topological structures of connectedness.
Figure \ref{fig:modular} illustrates the changes of network modular structure as parameters $p$ and $c$ vary. We used $\mu=1$ and $\sigma=0.25$ universally throughout study 2.

\begin{figure}[t]
\includegraphics[width=1\linewidth]{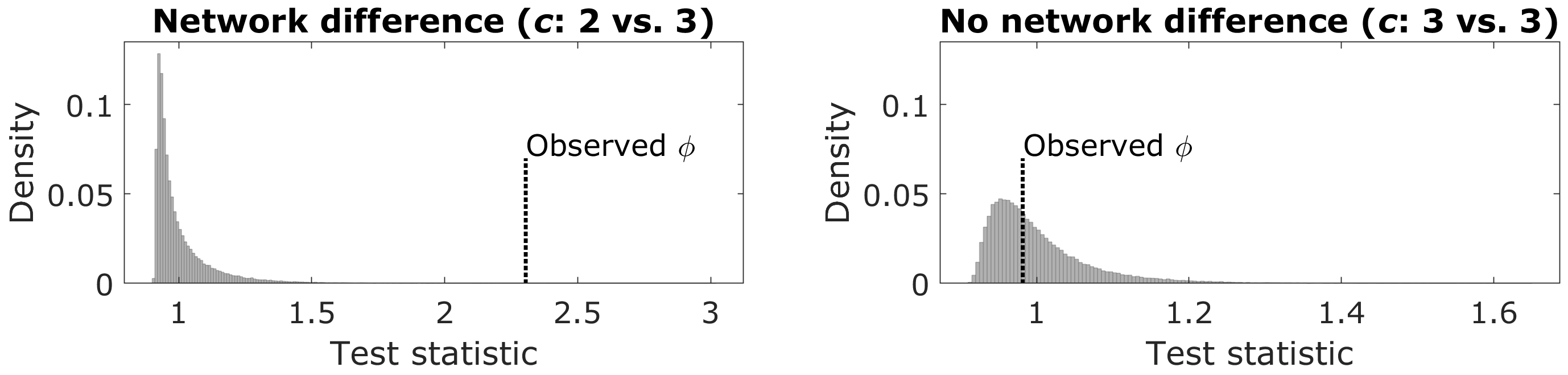}
\centering
\caption{Empirical distributions in Study 2. The empirical distribution of the ratio statistic $\phi_\mathcal{L}$ is generated by the permutation on two groups each consisting of 10 modular networks. Here we test if there is group difference in networks with  parameters $c=$ 2 vs. 3 (left) and 3 vs. 3 (right). As expected, the test based on the topological loss rejects the null hypothesis when there is group difference (left) while it does not reject when there is no difference (right).}
\label{fig:emp_dist}
\end{figure}

Based on the network model above, we simulated two groups of random modular networks $\mathcal{X} = (\mathcal{X}_1,\cdots,\mathcal{X}_m)$ and $\mathcal{Y} = (\mathcal{Y}_1,\cdots,\mathcal{Y}_n)$. If there is group difference, the topological loss is expected to be relatively small within groups and relatively large between groups. The average topological loss within the groups given by
\[
 \overline{\mathcal{L}}_{W} =\frac{\sum_{i < j} \mathcal{L}_{top}(\mathcal{X}_i,\mathcal{X}_j) + \sum_{i < j} \mathcal{L}_{top}(\mathcal{Y}_i,\mathcal{Y}_j)}{{m \choose 2} + {n \choose 2}} 
 \]
is expected to be smaller than the average topological loss between the groups given by
\[\overline{\mathcal{L}}_{B}=\frac{\sum_{i=1}^m \sum_{j=1}^n \mathcal{L}_{top}(\mathcal{X}_i,\mathcal{Y}_j)}{mn} .\]
We measure the disparity between groups as the  ratio $\phi_\mathcal{L}$
\[ \phi_\mathcal{L} = \overline{\mathcal{L}}_{B}\Big/ \overline{\mathcal{L}}_{W} .\]
If $\phi_\mathcal{L}$ is large, the groups differ significantly in network topology. On the other hand, if $\phi_\mathcal{L}$ is small, it is likely that there is no group difference. Similarly, we define the ratio statistic for graph matching cost $J$ as
\[ \phi_J = \overline{J}_{W}\Big/ \overline{J}_{B},\]
where $\overline{J}_{W}$ is the average graph matching cost within groups and $\overline{J}_{B}$ is the average graph matching cost between groups. Since the distributions of the ratios $\phi_\mathcal{L}$ and $\phi_{J}$ are unknown, the permutation test is used to determine the empirical distributions. Figure \ref{fig:emp_dist} displays the empirical  distribution of $\phi_\mathcal{L}$. By comparing the observed ratio $\phi_\mathcal{L}$ to the empirical distribution, we can determine the statistical significance of testing the group difference. However, when the sample size is large, existing matching algorithms are too slow for the permutation test. So we adapted a scalable online computation strategy as follows.

\begin{figure}[t]
\includegraphics[width=1\linewidth]{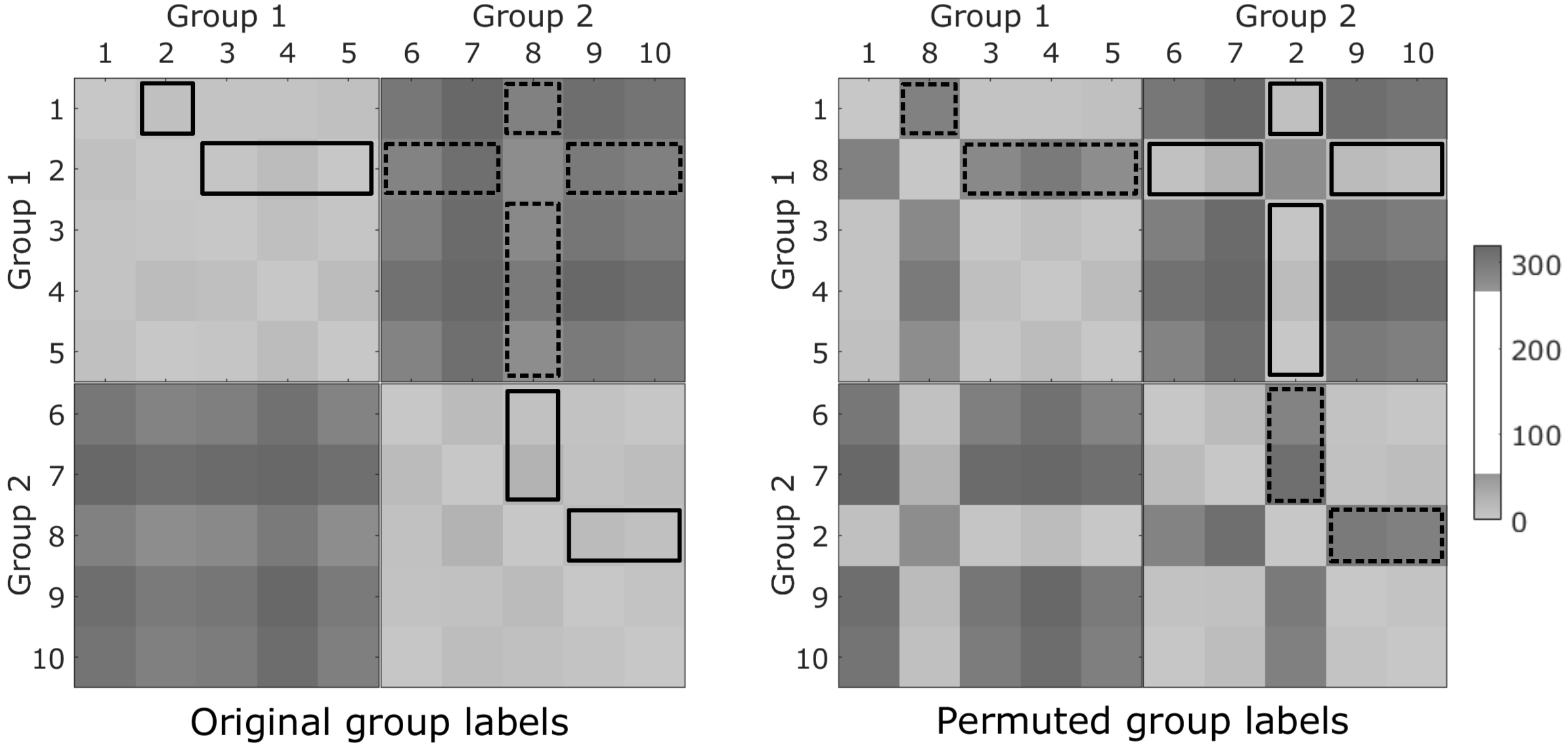}
\centering
\caption{Study 2. Two groups each consisting of five modular networks are simulated with parameters $c=$ 2 vs. 3.
Left: The $ij$-th entry represents the loss $\mathcal{L}_{top}$ between networks $i$ and $j$. The main diagonal consists of zeros since the topological loss between two identical networks is zero. Right: a transposition between the second network in Group 1 and the eighth network in Group 2. We do not need to recompute all the pairwise losses again but just rearrange the losses in the solid lines and dashed lines. Thus, we simply need to find out how the rearrangement changes the ratio statistic in an iterative manner. This enables us to efficiently perform the permutation test in an iterative fashion. We compute $\delta$ in equation (\ref{eq:transposition}) by subtracting the sum of entries within the solid lines from the sum of entries within the dashed lines.}
\label{fig:loss_mtx}
\end{figure}

Given two groups of networks, topological loss or graph matching cost for every pair of networks needs to be computed only once, which can then be arranged into a matrix whose rows and columns are networks. The $ij$-th entry is the loss between two networks corresponding to row $i$ and column $j$ (Figure \ref{fig:loss_mtx}). Once we obtain such a matrix, the permutation process is equivalent to rearranging rows and columns based on permuted group labels. There are $\frac{1}{2}{m+n \choose m}$ total number of permutations excluding the symmetry. Computing the ratio over a permutation requires re-summing over all such losses, which is time consuming. Instead, we performed the transposition procedure of swapping only one network per group and setting up iteration of how the ratio changes over the transposition.

\begin{figure}[t]
\includegraphics[width=1\linewidth]{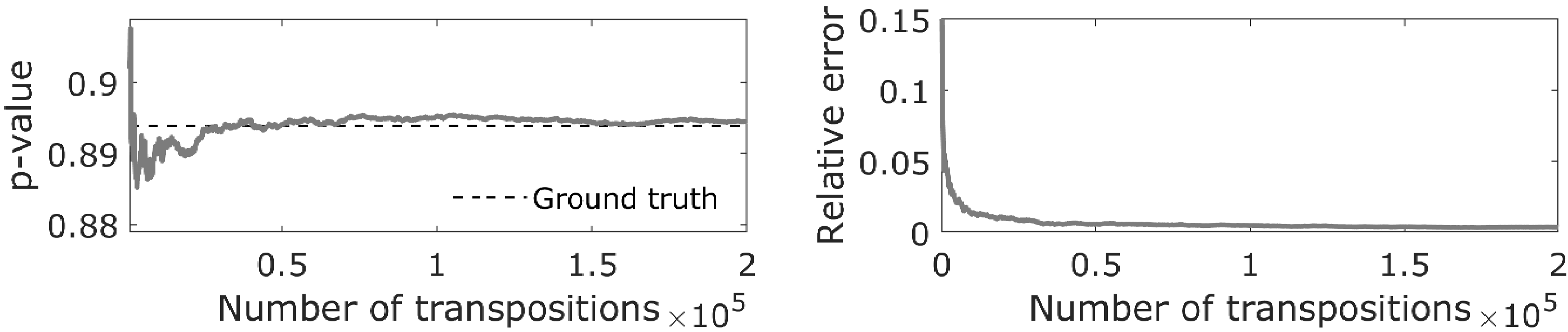}
\centering
\caption{The transposition test is used to determine the statistical significance of two groups each consisting of ten simulated networks. To further speed up the convergence rate, a random permutation is intermixed for every consecutive sequence of 500 transpositions. Left: the plot showing the convergence of $p$-value over 50,000 transpositions. For comparison, ground truth $p$-value is computed from the exact permutation test by enumerating every possible permutation exactly once. Right: the plot shows the average relative error against the ground truth across 100 independent simulations.}
\label{fig:trans_conv}
\end{figure}

We {\em transpose} $k$-th and $l$-th networks between the groups as
\begin{align*}
    \pi_{kl}(\mathcal{X}) &= (\mathcal{X}_1,\cdots,\mathcal{X}_{k-1},\mathcal{Y}_l,\mathcal{X}_{k+1},\cdots,\mathcal{X}_m), \\
    \pi_{kl}(\mathcal{Y}) &= (\mathcal{Y}_1,\cdots,\mathcal{Y}_{l-1},\mathcal{X}_k,\mathcal{Y}_{l+1},\cdots,\mathcal{Y}_n).
\end{align*}
Over transposition $\pi_{kl}$, the ratio statistic is changed from $\phi_L(\mathcal{X},\mathcal{Y})$ to $\phi_L\big(\pi_{kl}(\mathcal{X}),\pi_{kl}(\mathcal{Y})\big)$, which involves the following functions:
\begin{align*}
    \nu(\mathcal{X},\mathcal{Y}) &= \sum_{i < j} \mathcal{L}_{top}(\mathcal{X}_i,\mathcal{X}_j) + \sum_{i < j} \mathcal{L}_{top}(\mathcal{Y}_i,\mathcal{Y}_j), \\
    \omega(\mathcal{X},\mathcal{Y}) &= \sum_{i=1}^m \sum_{j=1}^n \mathcal{L}_{top}(\mathcal{X}_i,\mathcal{Y}_j),
\end{align*}
where  $\nu$ is the total sum of within-group losses and $\omega$ is the total sum of between-group losses.
Then we determine how $\nu$ and $\omega$ change over the transposition $\pi_{kl}$. As $\mathcal{X}_k$ and $\mathcal{Y}_l$ are swapped, the function $\nu$ is updated over the transposition $\pi_{kl}$ as (Figure \ref{fig:loss_mtx})
$$    \nu\big(\pi_{kl}(\mathcal{X}),\pi_{kl}(\mathcal{Y})\big) = \nu(\mathcal{X},\mathcal{Y}) + \delta(\mathcal{X},\mathcal{Y})$$
with
\begin{equation}
\delta(\mathcal{X},\mathcal{Y}) = \sum_{i \neq k} \mathcal{L}_{top}(\mathcal{Y}_l,\mathcal{X}_i) - \sum_{i \neq k} \mathcal{L}_{top}(\mathcal{X}_k,\mathcal{X}_i)  +  \sum_{i \neq l} \mathcal{L}_{top}(\mathcal{X}_k,\mathcal{Y}_i) - \sum_{i \neq l} \mathcal{L}_{top}(\mathcal{Y}_l,\mathcal{Y}_i).
\label{eq:transposition}
\end{equation}
Similarly, function $\omega$ is updated iteratively over the transposition $\pi_{kl}$ as
\begin{align*}
    \omega\big(\pi_{kl}(\mathcal{X}),\pi_{kl}(\mathcal{Y})\big) &= \omega(\mathcal{X},\mathcal{Y}) - \delta(\mathcal{X},\mathcal{Y}).
\end{align*}
The ratio statistic over the transposition is then computed as
\[\phi_{\mathcal{L}} \big(\pi_{kl}(\mathcal{X}),\pi_{kl}(\mathcal{Y})\big) = \frac{\omega\big(\pi_{kl}(\mathcal{X}),\pi_{kl}(\mathcal{Y})\big)}{\nu\big(\pi_{kl}(\mathcal{X}),\pi_{kl}(\mathcal{Y})\big)} \times \frac{{m \choose 2} + {n \choose 2}}{mn} .\]
For each transposition, we store information about function values $\nu$ and $\omega$, and update them sequentially. Each transposition requires manipulating $2(m+n-2)$ terms as opposed to ${m+n \choose 2}$ total number of terms over a random permutation. More transpositions than the number of permutations can be generated within the same amount of time, which speeds up the convergence  \citep{chung2019rapid}. To further accelerate the rate of convergence and avoid possible bias, we introduce a full permutation to the sequence of 500 consecutive transpositions. Figure \ref{fig:trans_conv} illustrates the convergence of transposition procedure.

In each simulation, we generated two groups each with 10 random modular networks. We then sequentially computed 200,000 random transpositions while interjecting a random permutation for every 500 transpositions and obtained the $p$-value. This indicates the convergence of $p$-value within 2 decimal places (within 0.01) in average. The simulations were independently performed 50 times and the average $p$-value was reported.

\paragraph*{Network difference} We compared two groups of networks generated by parameters $c=$ 2 vs. 3, 2 vs. 6 and 3 vs. 6 each with $d=12,18,24$ nodes and $p=0.6,0.8$ probability of connection within modules. Since each group exhibited a different modular structure, topological loss and graph matching costs were expected to detect the group difference. Table \ref{tab:gm_diff} summarizes the performance results. Networks with $d=12$ nodes might be too small to extract  distinct features used in each algorithm. Thus, all graph matching costs performed poorly while the topological loss performed reasonably well. When the number of nodes increases, all methods show overall performance improvement. In all settings, the topological loss significantly outperformed other graph matching algorithms.

\begin{table}[t] 
\caption{Study 2. Simulation study on network differences. The performance results are summarized as the average $p$-values for various parameter settings of $d$ (number of nodes), $c$ (number of modules) and $p$ (within-module probability).}
\label{tab:gm_diff}
\centering
\begin{tabular}{ccc|ccccc}
    \toprule
    $d$ & $c$ & $p$ & GA & SM & RRWM & IPFP & $\mathcal{L}_{top}$  \\
    \midrule 
    \multirow{6}{*}{12 vs. 12} & \multirow{2}{*}{2 vs. 3}   & 0.6 & $ 0.45 \pm 0.27 $ & $ 0.48 \pm 0.30 $ & $ 0.28 \pm 0.31 $ & $ 0.34 \pm 0.28 $ & $ 0.08 \pm 0.16 $ \\
                                 &                            & 0.8 & $ 0.26 \pm 0.24 $ & $ 0.30 \pm 0.28 $ & $ 0.06 \pm 0.12 $ & $ 0.28 \pm 0.28 $ & $ 0.01 \pm 0.03 $ \\
                                 & \multirow{2}{*}{2 vs. 6}   & 0.6 & $ 0.06 \pm 0.10 $ & $ 0.17 \pm 0.20 $ & $ 0.04 \pm 0.13 $ & $ 0.23 \pm 0.28 $ & $ 0.00 \pm 0.00 $ \\
                                 &                            & 0.8 & $ 0.00 \pm 0.01 $ & $ 0.01 \pm 0.03 $ & $ 0.00 \pm 0.00 $ & $ 0.02 \pm 0.04 $ & $ 0.00 \pm 0.00 $ \\
                                 & \multirow{2}{*}{3 vs. 6} & 0.6 & $ 0.40 \pm 0.29 $ & $ 0.35 \pm 0.28 $ & $ 0.24 \pm 0.26 $ & $ 0.35 \pm 0.28 $ & $ 0.06 \pm 0.13 $ \\
                                 &                            & 0.8 & $ 0.21 \pm 0.23 $ & $ 0.28 \pm 0.27 $ & $ 0.08 \pm 0.14 $ & $ 0.26 \pm 0.25 $ & $ 0.00 \pm 0.01 $ \\

    \bottomrule \toprule
    
    \multirow{6}{*}{18 vs. 18} & \multirow{2}{*}{2 vs. 3}   & 0.6 & $ 0.25 \pm 0.23 $ & $ 0.41 \pm 0.26 $ & $ 0.26 \pm 0.24 $ & $ 0.42 \pm 0.28 $ & $ 0.01 \pm 0.02 $ \\
                                 &                            & 0.8 & $ 0.12 \pm 0.17 $ & $ 0.19 \pm 0.22 $ & $ 0.00 \pm 0.00 $ & $ 0.04 \pm 0.05 $ & $ 0.00 \pm 0.00 $ \\
                                 & \multirow{2}{*}{2 vs. 6}   & 0.6 & $ 0.02 \pm 0.05 $ & $ 0.07 \pm 0.17 $ & $ 0.00 \pm 0.00 $ & $ 0.14 \pm 0.20 $ & $ 0.00 \pm 0.00 $ \\
                                 &                            & 0.8 & $ 0.00 \pm 0.00 $ & $ 0.00 \pm 0.00 $ & $ 0.00 \pm 0.00 $ & $ 0.00 \pm 0.00 $ & $ 0.00 \pm 0.00 $ \\
                                 & \multirow{2}{*}{3 vs. 6} & 0.6 & $ 0.28 \pm 0.24 $ & $ 0.37 \pm 0.31 $ & $ 0.21 \pm 0.24 $ & $ 0.37 \pm 0.30 $ & $ 0.01 \pm 0.01 $ \\
                                 &                            & 0.8 & $ 0.15 \pm 0.22 $ & $ 0.13 \pm 0.14 $ & $ 0.00 \pm 0.01 $ & $ 0.16 \pm 0.18 $ & $ 0.00 \pm 0.00 $ \\

    \bottomrule \toprule
    
    \multirow{6}{*}{24 vs. 24} & \multirow{2}{*}{2 vs. 3}   & 0.6 & $ 0.23 \pm 0.25 $ & $ 0.30 \pm 0.26 $ & $ 0.14 \pm 0.20 $ & $ 0.31 \pm 0.28 $ & $ 0.00 \pm 0.01 $ \\
                                 &                            & 0.8 & $ 0.06 \pm 0.11 $ & $ 0.12 \pm 0.19 $ & $ 0.00 \pm 0.00 $ & $ 0.01 \pm 0.05 $ & $ 0.00 \pm 0.00 $ \\
                                 & \multirow{2}{*}{2 vs. 6}   & 0.6 & $ 0.00 \pm 0.01 $ & $ 0.03 \pm 0.06 $ & $ 0.00 \pm 0.00 $ & $ 0.09 \pm 0.13 $ & $ 0.00 \pm 0.00 $ \\
                                 &                            & 0.8 & $ 0.00 \pm 0.00 $ & $ 0.00 \pm 0.00 $ & $ 0.00 \pm 0.00 $ & $ 0.00 \pm 0.00 $ & $ 0.00 \pm 0.00 $ \\
                                 & \multirow{2}{*}{3 vs. 6} & 0.6 & $ 0.24 \pm 0.26 $ & $ 0.29 \pm 0.28 $ & $ 0.10 \pm 0.13 $ & $ 0.37 \pm 0.26 $ & $ 0.00 \pm 0.00 $ \\
                                 &                            & 0.8 & $ 0.07 \pm 0.12 $ & $ 0.13 \pm 0.19 $ & $ 0.00 \pm 0.01 $ & $ 0.12 \pm 0.19 $ & $ 0.00 \pm 0.00 $ \\

    \bottomrule
    
\end{tabular}
\end{table}

\paragraph*{No network difference} We compared networks generated with parameters $c=$ 2 vs. 2, 3 vs. 3 and 6 vs. 6 each with $d=12,18,24$ nodes and $p=0.6,0.8$ probability of connection within modules. Since it was expected that there was no topological difference between networks generated using the same values for parameters $c,d$ and $p$, the topological loss and graph matching costs should not be able to detect the group difference. The performance result is summarized in Table \ref{tab:gm_nodiff}. All the methods performed relatively well when there was no group difference. The baseline graph matching methods have low sensitivity in detecting topological differences. They were unable to detect the network differences when the topological differences were subtle. 

While graph matching algorithms can be applied to dense networks, a small increment in the number of possible connections usually results in a combinatorial explosion of the amount of data to fit \citep{zhou2013deformable}. Thus, most high-order graph matching methods are often limited to sparse networks like binary trees. They are not practical in dense functional brain networks with a far large number of cycles. On the other hand, the proposed topological loss is able to detect such subtle topological differences without the added computational burden.

\begin{table}[t] 
\caption{Study 2. Simulation study on no network difference. The performance results are summarized as average $p$-values for various parameter settings of $d$ (number of nodes), $c$ (number of modules) and $p$ (within-module probability).}
\label{tab:gm_nodiff}
\centering
\begin{tabular}{ccc|ccccc}
    \toprule
    $d$ & $c$ & $p$ & GA & SM & RRWM & IPFP & $\mathcal{L}_{top}$ \\
    \midrule 
    \multirow{6}{*}{12 vs. 12} & \multirow{2}{*}{2 vs. 2}   & 0.6 & $ 0.49 \pm 0.27 $ & $ 0.46 \pm 0.30 $ & $ 0.51 \pm 0.30 $ & $ 0.47 \pm 0.28 $ & $ 0.53 \pm 0.29 $ \\
                                 &                            & 0.8 & $ 0.45 \pm 0.25 $ & $ 0.47 \pm 0.31 $ & $ 0.56 \pm 0.29 $ & $ 0.47 \pm 0.30 $ & $ 0.50 \pm 0.30 $ \\
                                 & \multirow{2}{*}{3 vs. 3}   & 0.6 & $ 0.45 \pm 0.32 $ & $ 0.44 \pm 0.26 $ & $ 0.47 \pm 0.27 $ & $ 0.51 \pm 0.30 $ & $ 0.46 \pm 0.31 $ \\
                                 &                            & 0.8 & $ 0.54 \pm 0.31 $ & $ 0.51 \pm 0.27 $ & $ 0.51 \pm 0.29 $ & $ 0.52 \pm 0.29 $ & $ 0.51 \pm 0.30 $ \\
                                 & \multirow{2}{*}{6 vs. 6} & 0.6 & $ 0.57 \pm 0.30 $ & $ 0.51 \pm 0.28 $ & $ 0.56 \pm 0.29 $ & $ 0.45 \pm 0.26 $ & $ 0.58 \pm 0.29 $ \\
                                 &                            & 0.8 & $ 0.55 \pm 0.29 $ & $ 0.48 \pm 0.26 $ & $ 0.52 \pm 0.27 $ & $ 0.54 \pm 0.30 $ & $ 0.49 \pm 0.27 $ \\

    \bottomrule \toprule
    
    \multirow{6}{*}{18 vs. 18} & \multirow{2}{*}{2 vs. 2}   & 0.6 & $ 0.48 \pm 0.26 $ & $ 0.49 \pm 0.32 $ & $ 0.54 \pm 0.29 $ & $ 0.47 \pm 0.30 $ & $ 0.54 \pm 0.31 $ \\
                                 &                            & 0.8 & $ 0.52 \pm 0.28 $ & $ 0.50 \pm 0.28 $ & $ 0.46 \pm 0.30 $ & $ 0.52 \pm 0.25 $ & $ 0.50 \pm 0.26 $ \\
                                 & \multirow{2}{*}{3 vs. 3}   & 0.6 & $ 0.49 \pm 0.28 $ & $ 0.58 \pm 0.31 $ & $ 0.43 \pm 0.28 $ & $ 0.51 \pm 0.27 $ & $ 0.53 \pm 0.30 $ \\
                                 &                            & 0.8 & $ 0.46 \pm 0.30 $ & $ 0.51 \pm 0.27 $ & $ 0.52 \pm 0.33 $ & $ 0.45 \pm 0.29 $ & $ 0.53 \pm 0.27 $ \\
                                 & \multirow{2}{*}{6 vs. 6} & 0.6 & $ 0.53 \pm 0.28 $ & $ 0.48 \pm 0.30 $ & $ 0.51 \pm 0.30 $ & $ 0.45 \pm 0.29 $ & $ 0.44 \pm 0.33 $ \\
                                 &                            & 0.8 & $ 0.54 \pm 0.27 $ & $ 0.52 \pm 0.30 $ & $ 0.48 \pm 0.26 $ & $ 0.52 \pm 0.31 $ & $ 0.43 \pm 0.30 $ \\

    \bottomrule \toprule
    
    \multirow{6}{*}{24 vs. 24} & \multirow{2}{*}{2 vs. 2}   & 0.6 & $ 0.52 \pm 0.28 $ & $ 0.49 \pm 0.30 $ & $ 0.50 \pm 0.30 $ & $ 0.48 \pm 0.28 $ & $ 0.55 \pm 0.26 $ \\
                                 &                            & 0.8 & $ 0.53 \pm 0.27 $ & $ 0.56 \pm 0.30 $ & $ 0.51 \pm 0.30 $ & $ 0.56 \pm 0.32 $ & $ 0.52 \pm 0.30 $ \\
                                 & \multirow{2}{*}{3 vs. 3}   & 0.6 & $ 0.48 \pm 0.29 $ & $ 0.54 \pm 0.27 $ & $ 0.49 \pm 0.26 $ & $ 0.49 \pm 0.30 $ & $ 0.52 \pm 0.30 $ \\
                                 &                            & 0.8 & $ 0.55 \pm 0.29 $ & $ 0.49 \pm 0.27 $ & $ 0.52 \pm 0.28 $ & $ 0.49 \pm 0.30 $ & $ 0.47 \pm 0.26 $ \\
                                 & \multirow{2}{*}{6 vs. 6} & 0.6 & $ 0.47 \pm 0.30 $ & $ 0.45 \pm 0.31 $ & $ 0.51 \pm 0.29 $ & $ 0.56 \pm 0.28 $ & $ 0.49 \pm 0.29 $ \\
                                 &                            & 0.8 & $ 0.51 \pm 0.30 $ & $ 0.47 \pm 0.28 $ & $ 0.54 \pm 0.28 $ & $ 0.56 \pm 0.31 $ & $ 0.51 \pm 0.31 $ \\

    \bottomrule
    
\end{tabular}
\end{table}

\section{Application}

In standard brain network modeling framework, the whole brain is parcellated into $d$ disjoint regions, where $d$ is  usually a few hundreds \citep{arslan.2018,desikan.2006,eickhoff.2018,fan.2016,fornito.2010,fornito.2016,glasser.2016,gong.2009,hagmann.2007,schaefer.2017,shattuck.2008,tzourio.2002,zalesky.2010}.  Subsequently, functional or structural information is overlaid on top of the parcellation to obtain $d \times d$ connectivity matrices that measure the strength of connectivity between brain regions (Figure \ref{fig:schematic_toplearning}). These disjoint brain regions form nodes in the brain network. Connectivity between brain regions that defines edges in the brain network is usually determined by the type of imaging modality \citep{ombao2016handbook}. Structural connectivity is obtained through diffusion MRI (dMRI), which traces the white matter fibers connecting brain regions. The strength of structural connectivity between brain regions is determined by the number of fibers passing through them \citep{fornito.2016}. The structural brain network is expected to exhibit sparse topology without many loops or cycles (Figure \ref{fig:schematic_toplearning}) \citep{chung.2011.SPIE,gong.2009,zhang.2018}. On the other hand, functional connectivity obtained from the resting-state functional MRI (fMRI) is often computed  as the Pearson correlation coefficient between brain regions \citep{bryant2017lcn,shappell.2019}. While structural connectivity provides information whether the brain regions are physically connected through white matter fibers, functional connectivity exhibits connections  between two regions without the direct neuroanatomical connections  through additional intermediate connections \citep{honey2007network,honey2009predicting}. Thus, resting-state functional brain networks are often very dense with thousands of cycles. Both structural and functional brain networks provide topologically different information. Existing graph theory based brain network analyses have shown that there is some common topological profile that is conserved for both structural and functional brain networks \citep{bullmore2009complex}. However, due to the difficulty of integrating both networks in a coherent statistical framework, not much research has been done on integrating such networks at the localized connection level. Many previous studies focus on comparing summary graph theory features across different networks \citep{bullmore2009complex,ginestet2011brain,karas2019brain}. Although there are few studies that focus on fusing networks derived from both modalities probabilistically, such methods can easily destroy the aforementioned topological difference of the networks \citep{kang.2017,xue2015multimodal}. This motivates the development of a model for multimodal networks that can integrate networks of different topology at the localized connection level.

\subsection{Dataset and pre-processing} 

Dataset studied in this paper is the resting-state fMRI of 412 subjects collected as part of the Human Connectome Project (HCP) twin study \citep{van2012human,van2013wu}. The fMRI were collected over 14 minutes and 33 seconds using a gradient-echo-planar imaging (EPI) sequence with multiband factor 8, repetition time (TR) 720 ms, time echo (TE) 33.1 ms, flip angle $52^\circ$, $104\times90$ (RO$\times$PE) matrix size, 72 slices, 2 mm isotropic voxels, and 1200 time points. Subjects without fMRI or full 1200 time points were excluded. During scanning, participants were at rest with eyes open with relaxed fixation on a projected bright cross-hair on a dark background  \citep{van2013wu}. The standard minimal pre-processing pipelines \citep{glasser2013minimal} were applied on the fMRI scans including spatial  distortion removal \citep{andersson2003correct,jovicich2006reliability}, motion correction \citep{jenkinson2001global,jenkinson2002improved}, bias field reduction \citep{glasser.2011}, registration to the structural MNI template, and data masking using the brain mask obtained from FreeSurfer \citep{glasser2013minimal}. This resulted in the resting-state functional time series with $91\times 109 \times 91$ 2-mm isotropic voxels at 1200 time points.
The subject ranged from 22 to 36 years in age with average age $29.24 \pm 3.39$ years. There are 172 males and 240 females. Among them, there are 131 monozygotic (MZ) twin pairs and 75 same-sex dizygotic (DZ) twin pairs. 

Subsequently,  we employed the Automated Anatomical Labeling (AAL) template to parcellate the brain volume into 116 non-overlapping anatomical regions \citep{tzourio.2002}  (Figure \ref{fig:schematic_toplearning}). We averaged fMRI across voxels within each brain parcellation, resulting in 116 average fMRI time series with 1200 time points for each subject.
Previous studies reported that  head movement produces spatial artifacts in functional connectivity \citep{caballero.2017,power.2012,satterthwaite.2012,van.2012}. Thus, we scrubbed the data with significant head motion using the framewise displacement (FD) from the three translational displacements and three rotational displacements at each time point  to measure  the head movement from one volume to the next \citep{huang.2020.NM,power.2012,van.2012}. We excluded 12 subjects having excessive head movement, resulting in fMRI data of 400 subjects (168 males and 232 females). Among the remaining 400 subjects, there are $p=124$ MZ twin pairs and $q=70$ same-sex DZ twin pairs. The first 20 time points were removed from all subjects to avoid artifacts in the fMRI data, leaving 1180 time points per subject \citep{diedrichsen.2005,shah.2016}.

For dMRI, we use the template structural brain network $P$ (Figure \ref{fig:schematic_toplearning}) published in previous studies \citep{chung2019rapid, song.2021.MICCAI}. Pre-processing pipeline for $P$ reported in those studies is as follows.
The white matter fiber orientation information was extracted by multi-shell, multi-tissue constrained spherical deconvolution from different tissue types such as white matter and gray matter \citep{callaghan.1988,jeurissen.2014}. The fiber orientation distribution functions were estimated and apparent fiber densities were exploited to produce the reliable white and gray matter volume maps \citep{christiaens.2015,jeurissen.2014}. Subsequently, multiple random seeds were selected in each voxel to generate about 10 million initial streamlines per subject with the maximum fiber tract length at 250 mm and FA larger than 0.06 using MRtrix3 \citep{tournier.2012,xie.2018}. The Spherical-Deconvolution Informed Filtering of Tractograms (SIFT2) technique making use of complete streamlines was subsequently applied to generate more biologically accurate brain connectivity, which results in about 1 million tracts per subject \citep{smith.2015}. Nonlinear diffeomorphic registration between subject images to the template was performed using ANTS \citep{avants.2008,avants.2011}. AAL was used to parcellate the brain into 116 regions \citep{tzourio.2002}. The subject-level connectivity matrices were constructed by counting the number of tracts connecting between brain regions. 
The structural brain network $P$ that serves as the template where all the functional networks are aligned was obtained by computing the one sample $t$-statistic map $P$ over all the subjects and rescaling $t$-statistics between 0 to 2 using the hyperbolic tangent function $\tanh$, then adding 1.

\subsection{Learning individual networks}

For subject $k$, we have resting-state fMRI time series $x=(x_1,x_2,...,x_{1180})$ for region $i$ and $y=(y_1,y_2,...,y_{1180})$ for region $j$ with 1180 time points. Correlation $\rho^k_{ij}$ between regions $i$ and $j$ is computed using the Pearson correlation between $x$ and $y$. This gives the correlation matrix $C_k = (\rho^k_{ij})$, which is used as the baseline against our proposed method. We then translate and scale the correlation as $w_{ij}^k = \sqrt{(1-\rho^k_{ij})\big/2}$, which is a metric  \citep{chung2019exact}.  The subject-level functional brain network is given by $G_k = (V, w^k)$.

We apply the topological learning to estimate the subject-level network $\Theta_k(\lambda_k)$ by minimizing the objective function (\ref{eq:training-sub}) using the individual network $G_k$ and the structural network $P$:
\[\widehat \Theta_k(\lambda_k) =
 \argmin_\Theta \mathcal{L}_F(\Theta,G_k) + \lambda_k \mathcal{L}_{top}(\Theta,P) .
\]
$\Theta$ is initialized to $G_k$. $\lambda_k$ was found to be $1.0000 \pm 0.0002$ and thus we globally used $\lambda=1.0000$ (Figure \ref{fig:loss_vs_lambda_plot}).

\subsection{Heritability in twins}
We employ our method to explore which regions of brain networks are genetically heritable. In particular, we determine if the estimated network $\widehat\Theta_k$ is genetically heritable in twins. At each edge, let $(a_{1l},a_{2l})$ be the $l$-th twin pair in MZ-twin and $(b_{1l},b_{2l})$ be the $l$-th twin pair in DZ-twin. 
MZ-twin and DZ-twin pairs are then represented as
\[a=
\begin{pmatrix}
a_{11} & \cdots & a_{1p} \\
a_{21} & \cdots & a_{2p}
\end{pmatrix}, \quad
b=
\begin{pmatrix}
b_{11} & \cdots & b_{1q} \\
b_{21} & \cdots & b_{2q}
\end{pmatrix}
.
\]
Let $a_r=(a_{r1},a_{r2},...,a_{rp})$ and $b_r=(b_{r1},b_{r2},...,b_{rq})$ be the $r$-th rows. Then MZ-correlation is computed as the Pearson correlation $\gamma^{MZ}(a_1,a_2)$ between $a_1$ and $a_2$. Similarly, DZ-correlation $\gamma^{DZ}(b_1,b_2)$ is computed. 
In well established ACE genetic model, the heritability index (HI) $h$, which determines the amount of variation caused by genetic factors in population, is estimated using Falconer's formula \citep{falconer.1995}:
$$ h(a,b) = 2(\gamma^{MZ} - \gamma^{DZ}) .$$
Since the order of the twins is interchangeable, we can {\em transpose} the $l$-th twin pair in MZ-twin as
\begin{align*}
\pi_l(a_1) &= (a_{11},...,a_{1,l-1},a_{2l},a_{1,l+1},...,a_{1p}), \\
\pi_l(a_2) &= (a_{21},...,a_{2,l-1},a_{1l},a_{2,l+1},...,a_{2p})
\end{align*}
and obtain another MZ-correlation $\gamma^{MZ}(\pi_l(a_1),\pi_l(a_2))$. Likewise, we can obtain many different correlation values for DZ-twin.
To this end, we perform a sequence of random transpositions iteratively to estimate the twin correlations $\gamma^{MZ}$ and $\gamma^{DZ}$ sequentially, similar to the transposition test used in the simulation study 2, as follows.

\sloppy Over the transposition $\pi_l$, the MZ-correlation is changed from $\gamma^{MZ}(a_1,a_2)$ to $\gamma^{MZ}(\pi_l(a_1),\pi_l(a_2))$, which involves the following functions:
\begin{align*}
\nu(a_r) &= \sum_{j=1}^p a_{rj}, \\
\omega(a_r,a_s) &= \sum_{j=1}^p (a_{rj} - \nu(a_r)\big/p)(a_{sj} - \nu(a_s)\big/p) .
\end{align*}
The functions $\nu$ and $\omega$ are updated iteratively over the transposition $\pi_l$ as
\begin{align*}
\nu(\pi_l(a_r)) &= \nu(a_r) - a_{rl} +a_{sl}, \\
\omega(\pi_l(a_r),\pi_l(a_s)) &= \omega(a_r,a_s) + (a_{rl} - a_{sl})^2\big/p - (a_{rl} - a_{sl})(\mu(a_r) - \mu(a_s))\big/p .
\end{align*}
Then MZ-correlation after the transposition is calculated as
\[ \gamma^{MZ}(\pi_l(a_1),\pi_l(a_2)) = \frac{\omega(\pi_l(a_1),\pi_l(a_2))}{\sqrt{\omega(\pi_l(a_1),\pi_l(a_1))\omega(\pi_l(a_2),\pi_l(a_2))}} .\]
The time complexity for computing correlation iteratively is 33 operations per transposition, which is significantly more efficient than that of direct correlation computation per permutation. In our numerical implementation, we sequentially perform random transpositions $\pi_{l_1},\pi_{l_2},...,\pi_{l_J}$, which result in $J$ different twin correlations. Let
\[ \kappa_1 = \pi_{l_1},\quad \kappa_2 = \pi_{l_2} \circ \pi_{l_1},\quad \cdots,\quad \kappa_J = \pi_{l_J} \circ \cdots \circ \pi_{l_2} \circ \pi_{l_1}\]
be the sequence of transpositions. The average MZ-correlation $\bar \gamma^{MZ}_J$  is then given by
\[\bar \gamma^{MZ}_J = \frac{1}{J} \sum_{j=1}^J \gamma^{MZ}(\kappa_j(a_1),\kappa_j(a_2)) ,\]
which is iteratively updated as
\[\bar \gamma^{MZ}_J = \frac{J-1}{J} \bar\gamma^{MZ}_{J-1} + \frac{1}{J}\gamma^{MZ}(\kappa_J(a_1),\kappa_J(a_2)) .\]
The average correlation $\bar \gamma^{MZ}_J$ converges to the true underlying twin correlation $\gamma^{MZ}$ for sufficiently large $J$. Similarly, DZ-correlation $\gamma^{DZ}$ is estimated.

\subsection{Results}

\begin{figure}[t]
    \centering
    \includegraphics[width=1\textwidth]{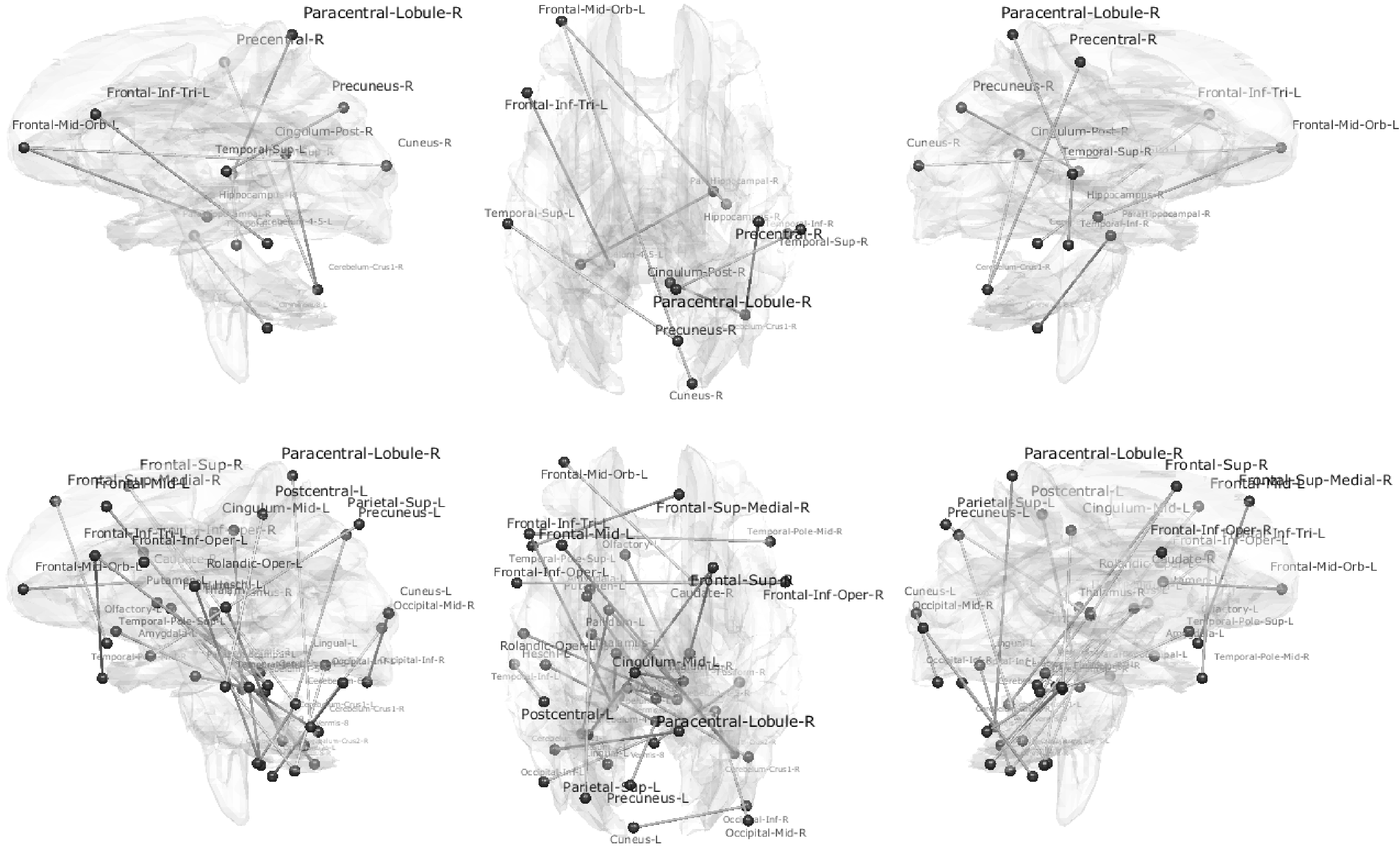}
    \caption{Most heritable connections above HI $\ge 1$ are shown using the Pearson correlation baseline (top row) and our topologically learned network approach (bottom row).}
    \label{fig:twincorr_HI}
\end{figure}

\begin{figure}[t]
    \centering
    \includegraphics[width=1\textwidth]{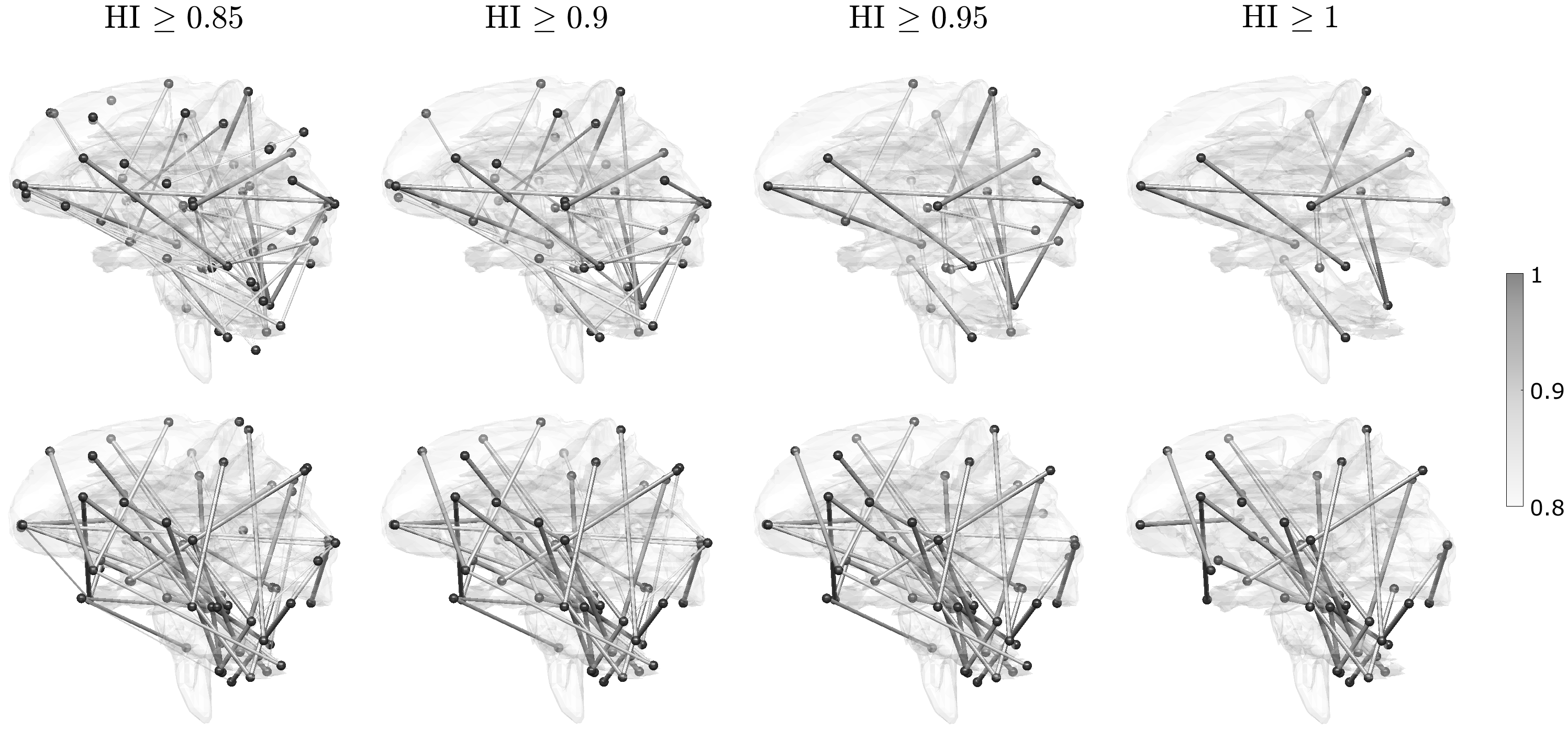}
    \caption{HI-maps for the Pearson correlation baseline (top row) and our topologically learned network approach (bottom row) are thresholded at different HI $\geq 0.85, 0.9, 0.95$ and 1.}
    \label{fig:twincorr_HI_dynamic}
\end{figure}

Using the transposition method, we randomly transposed twins and updated the correlations for $50,000$ times. This process was repeated $100$ times and the total $50,000 \times 100$ correlations were used to estimate the underlying MZ- and DZ-correlations. At each edge, the standard deviation of the average correlations from 100 results was smaller than 0.01, which indicates the convergence of the estimate within two decimal places in average.

We computed HI-maps using the original correlation matrices $C_k$ and the proposed topologically learned networks $\widehat \Theta_k$. Figures \ref{fig:twincorr_HI} displays most heritable connections with 100\% heritability, indicating that the topologically learned networks $\widehat \Theta_k$ show more connections than the original Pearson correlation matrices $C_k$. Table \ref{tab:HItable} shows that left superior parietal lobule and left amygdala connection has the strongest heritability among many other connections using the topologically learned networks.
Since the networks $\widehat \Theta_k$ inherited sparse topology from the template structural brain network $P$ (Figure \ref{fig:cycles}), short-lived cycles in the functional networks were expected to be removed, resulting in the increased statistical sensitivity to subtle genetic signals.
There are significant connection overlaps between the standard method and the topology-based approach, but our approach detects more connections with higher HI. 
Figure \ref{fig:twincorr_HI_dynamic} displays heritable connections above HI value 0.85.
If correlation thresholds are altered to values below 0.85, more connections will appear, resulting in more overlaps.

\begin{table}[t]
\caption{Top ten most heritable connections with HI $\ge 1$ using the proposed topologically learned networks.}
\label{tab:HItable}
\centering
\small
\begin{tabular}{l}
\toprule
{\bf Most heritable connections} \\ 
\midrule 
Left superior parietal lobule \;--\; Left amygdala \\
Left lobule VIIIB of cerebellar hemisphere \;--\; Left globus pallidus \\
Left lobule III of cerebellar hemisphere \;--\; Right crus I of cerebellar hemisphere \\
Left \;--\; Right opercular part of inferior frontal gyrus \\
Left lobule IV, V of cerebellar hemisphere \;--\; Left thalamus \\
Left lobule IX of cerebellar hemisphere \;--\; Left lobule VI of cerebellar hemisphere \\
Right thalamus \;--\; Right superior frontal gyrus, dorsolateral \\
Left middle frontal gyrus, orbital part \;--\; Right caudate nucleus \\
Right crus II of cerebellar hemisphere \;--\; Left globus pallidus \\
Lobule VIII of vermis \;--\; Right fusiform gyrus \\ 
\bottomrule  
\end{tabular}
\end{table}

\section{Discussion}

We present a new topological loss for graphs that provides the optimal matching and alignment at the edge level. Unlike many existing graph matching algorithms \citep{babai.1983,guo2020representations,tian2012convergence,wang2020learning,yu2018generalizing,zhang2019kergm,zhou2013deformable} that provide matching costs without explicit identification of how edges are matched to each other, the proposed method identifies the edge-to-edge correspondence explicitly using the birth-death decomposition. Such explicit mapping enables us to develop the subsequent topological learning framework that can integrate networks of different sizes and topology. Due to the wide availability of various network data including social networks, computer networks, artificial networks such as convolutional neural networks, our method can be easily adapted for other network applications where matching of whole networks or subnetworks is needed.

The limitation of the topological loss is the inability to discriminate geometrically different networks that have identical topology. We can obtain topologically identical networks by mirror reflection. Since the human brain is asymmetric across hemispheres \citep{toga.2003}, the ability to discriminate such networks is critical. In our application, we introduced the Frobenius loss to geometrically constrain brain networks. We provided one possible approach for combining the topological and geometrical losses together. Hopefully this paper serves as the springboard for more refined models in the future.

Among many different learning tasks, the proposed method is illustrated with averaging and regression. Our method can be used to average networks of different sizes and topology, which is challenging using prior methods. 
Our method is further used to set up the optimization-based regression models at the subject and group levels. We believe the proposed method can be easily adapted to other types of network learning tasks. 
For example, it is well known that $k$-means clustering does not perform well against more geometry based clustering methods such as spectral clustering \citep{kriegel2009clustering,ng.2002}. 
A new clustering method that utilizes meaningful network topology is expected to outperform $k$-means and spectral clustering methods.
Recently, it was shown that topological clustering using our proposed framework demonstrates very strong performance for clustering measured functional brain networks used to evaluate biomarkers of the neural basis of consciousness \citep{songdechakraiwut2022fast}. We believe that the demonstrated effectiveness and computational elegance of our topological framework will have a high impact on the analysis of brain networks.

Existing network predictive models typically employ various forms of regressions such as linear models and logistic regressions that incorporate the accumulated effect of network features into prediction scores \citep{arslan.2018,eickhoff.2016,goodfellow.2016,huang.2021.NCA,kong.2019,rottschy.2012, zhang.2019}. While the use of these methods may be reasonble to discover the underlying network features and their combinations for prediction, the development of new methods using the topological loss may provide additional insights into network analyses. This is left as a future study.

The proposed method is applied to multimodal brain networks in the twin brain imaging study. We determine the extent to which brain networks are genetically heritable by using the heritability index, which is twice the difference between MZ- and DZ-twin correlations. Due to the possible swapping of twin pairing, twin correlations are not unique. This has been considered as a main weakness of the widely used ACE model in genetics. We remedy the problem by computing a sufficiently large number of permutations over twin labels through the transposition test. This enables us to perform the network analysis at the edge level even when network shapes and topologies are different. We believe that the transposition test would be useful in various resampling problems beyond twin correlations. This is left as a future study.

%
%

\begin{acks}[Acknowledgments]
We thank Hernando Ombao of King Abdullah University of Science and Technology (KAUST) for discussion on brain network modeling. We thank Gary Shiu of University of Wisconsin--Madison for discussion on stability theorem. We thank Li Shen of University of Pennsylvania for providing the template structural brain network used in our twin brain imaging study. We thank Shih-Gu Huang of National University of Singapore for providing fMRI pre-processing support.
\end{acks}

%
\begin{funding}
This study is funded by NIH R01 EB022856, EB02875, NSF MDS-2010778.
\end{funding}



\bibliographystyle{imsart-nameyear} 
\bibliography{reference1,reference2}       


\end{document}